\documentclass[prd,aps,floats,floatfix,eqsecnum,nofootinbib]{revtex4}
\usepackage{verbatim,graphicx,amssymb,amsbsy,bm,amsmath,rotating,epsfig}
\usepackage{psfrag}
\usepackage{hyperref}
\newcommand{\be}{\begin{equation}}
\newcommand{\ee}{\end{equation}}
\newcommand{\bea}{\begin{eqnarray}}
\newcommand{\eea}{\end{eqnarray}}

\newcommand{\vx}{\ensuremath{\vec{x}}}

\newcommand{\epv}{\epsilon_v}
\begin{document}
\title{The pre-inflationary and inflationary fast-roll eras and their signatures
in the low CMB multipoles}
\author{\bf C. Destri $^{(a)}$} \email{Claudio.Destri@mib.infn.it}
\author{\bf H. J. de Vega $^{(b,c)}$}
\email{devega@lpthe.jussieu.fr} 
\author{\bf N. G. Sanchez $^{(c)}$}
\email{Norma.Sanchez@obspm.fr} 
\affiliation{$^{(a)}$ Dipartimento di Fisica G. Occhialini, Universit\`a
Milano-Bicocca and INFN, sezione di Milano-Bicocca, Piazza della Scienza 3,
20126 Milano, Italia. \\
$^{(b)}$ LPTHE, Universit\'e
Pierre et Marie Curie (Paris VI) et Denis Diderot (Paris VII),
Laboratoire Associ\'e au CNRS UMR 7589, Tour 24, 5\`eme. \'etage, 
Boite 126, 4, Place Jussieu, 75252 Paris, Cedex 05, France. \\
$^{(c)}$ Observatoire de Paris,
LERMA. Laboratoire Associ\'e au CNRS UMR 8112.
 \\61, Avenue de l'Observatoire, 75014 Paris, France.}
\date{\today}
\begin{abstract}
We study the entire coupled evolution of the inflaton $ \phi(t) $ and 
the scale factor $ a(t) $ for general initial conditions
$ \phi(t_0) $ and $ d\phi(t_0)/dt $ at a given initial time $ t_0 $. 
The {\bf generic} early universe evolution has three stages:  decelerated fast-roll 
followed by inflationary fast-roll and then inflationary slow-roll 
(an attractor always reached for {\bf generic} 
initial conditions). This evolution is valid for all regular inflaton potentials 
$ v(\phi) $. In addition, we find a special (extreme) slow-roll 
solution starting at $ t = -\infty $ in which the fast-roll stages are absent. 
At some time $ t = t_* $, the evolution backwards in time from 
$ t_0 $ reaches generically a mathematical singularity
where $ a(t) $ vanishes and the Hubble parameter becomes singular. 
We determine the general behaviour near the singularity.
The classical homogeneous inflaton description turns to be valid for 
$ t-t_* > 10 \; t_{Planck} $ well before the beginning of inflation,
quantum loop effects are negligible there. The singularity 
is never reached in the validity region of the classical treatment and therefore
it {\bf is not a real physical phenomenon} here.
Fast-roll and slow-roll regimes are analyzed in detail including the equation 
of state evolution, both analytically and numerically. 
The characteristic time scale of the fast-roll era turns to be 
$ t_1= (1/m) \; \sqrt{V(0)/[3 \; M^4] } \sim 10^4 \;  t_{Planck} $ where 
$ V $ is the double-well inflaton potential, $ m $ is the inflaton mass and $ M $ the energy 
scale of inflation. The {\bf whole} evolution of the fluctuations along the 
decelerated and inflationary 
fast-roll and slow-roll eras is computed. The Bunch-Davies initial conditions 
(BDic) are generalized for the present case in which the potential felt by the 
fluctuations can never be neglected. The fluctuations feel 
a {\bf singular attractive} potential near the $ t = t_* $ singularity 
(as in the case of a particle in a central singular potential) with {\bf exactly} the 
{\bf critical} strength ($ -1/4 $) allowing the fall to the centre.
Precisely, the fluctuations exhibit logarithmic behaviour describing the fall to 
$ t = t_* $. The power spectrum gets dynamically modified by the effect of the 
fast-roll eras and the choice of BDic at a finite time 
through the transfer function $ D(k) $ of initial conditions. The power spectrum
vanishes at $ k = 0 . \; D(k) $ presents a first peak for $ k \sim 2/\eta_0 $
($ \eta_0 $ being the conformal initial time), 
then oscillates with decreasing amplitude and vanishes asymptotically for $ k \to \infty $.
The transfer function $ D(k) $ affects the {\bf low} CMB multipoles $ C_{\ell} $:
the change $ \Delta C_{\ell}/ C_{\ell} $ for $ 1 \leq \ell \leq 5 $ is computed as a
function of the starting instant of the fluctuations $ t_0 $.
CMB quadrupole observations indicate large {\bf suppressions} which
are well reproduced for the range  $ t_0 - t_\ast \gtrsim 0.05/m \simeq 10100 \; t_{Planck} $.
\end{abstract}
\maketitle
\tableofcontents

\section{Introduction and summary of results}

Since the Universe expands exponentially fast during inflation, gradients are 
exponentially erased and can be neglected. At the same time, the exponential 
stretching of spatial lengths 
{\bf classicalizes} the physics and allows a {\bf classical} treatment. 
One can therefore consider a homogeneous and classical inflaton field which thus 
determines self-consistently a homogenous and isotropic Friedman-Robertson Walker 
metric sourced by this inflaton.

This treatment is valid for early times well after the Planck time
$ t = 10^{-44} $ sec., at which the quantum fluctuations are expected to be large and
thus a full quantum gravity treatment is required. 

\medskip

In this paper we study the entire coupled evolution of the inflaton field $ \phi(t) $ and 
the scale factor $ a(t) $ of the metric for generic initial conditions,  
fixed by the values of $ \phi(t_0) $ and $ d\phi(t_0)/dt $ at a given 
initial time $ t_0 $. 

\medskip

We show that the {\bf generic} early universe evolution has three stages: a 
decelerated fast-roll stage followed by an inflationary fast-roll stage and then 
by a slow-roll inflationary regime which is an attractor always reached for 
{\bf generic} initial conditions. This evolution is valid for all regular inflaton potentials.
In addition, we find a particular (extreme) 
slow-roll solution starting from $ t = -\infty $ in which the fast-roll stages 
are absent. 

\medskip

The evolution backwards in time from 
$ t_0 $ reachs  generically a mathematical singularity at some time $ t = t_* $ 
where the scale factor $ a(t) $ vanishes, and the Hubble parameter becomes singular. 

We find the general behaviour of the inflaton and the scale factor 
near the singularity as given by eqs. (\ref{sing1})-(\ref{asegs}) and determine the 
validity of the classical approximation, namely $ (H/M_{Pl})^2 \ll 1 $. 
It must be stressed that such mathematical singularity is attained extrapolating 
the classical treatment where it is {\bf no more valid}. The singularity 
is never reached in the validity region of the classical treatment and therefore
such mathematical singularity {\bf is not a real physical phenomenon} here.

Quantum loops effects turns to be less than 1\% for $ t - t_* > 10^{-42} $ sec and therefore
the classical treatment of the inflaton and the space-time can be trusted
well before the begining of inflation.

\medskip

The fast-roll (both decelerated and inflationary) and slow-roll regimes are analyzed in 
detail, with both the exact numerical evolution and an analytic approximation, 
and the whole equation of state evolution in the three regimes. 
We consider here the double well (broken symmetric) fourth order inflaton potential
since it gives the best description of the CMB+LSS data \cite{mcmc,biblia}
within the Ginsburg-Landau effective theory approach we follow.

\medskip

The characteristic time scale of the fast-roll era turns to be 
$ t_1= (1/m) \; \sqrt{V(0)/[3 \; M^4] } \sim 10^4 \;  t_{Planck} $ where 
$ V(0) $ is the double well inflaton potential at zero inflaton field,
$ m $ is the inflaton mass and $ M $ the energy scale of inflation.
The time scale of the inflaton in the extreme slow roll solution goes as 
the inverse of $ t_1 $, namely $ 1/[m^2 \; t_1] $.

\medskip

We study the {\bf whole} evolution of the curvature and tensor fluctuations along 
the three succesive regimes: decelerated fast-roll followed by inflationary 
fast-roll and then inflationary slow-roll, and compute the power spectrum
by the end of inflation. 
The fluctuations feel a {\bf singular attractive} potential
near the $ t = t_* $ singularity 
(as in the case of a particle in a central singular potential) with {\bf exactly} the
{\bf critical} strength ($ -1/4 $) for which the fall to the centre becomes possible.
Precisely, the logarithmic behaviour of the fluctuations
for $ t \to t_* $  eq.(\ref{sr0}) describes the fall to $ t = t_* $ for the 
critical strength of the potential $ W_\mathcal{R} $ felt by the fluctuations. 

\medskip

We generalize the Bunch-Davies initial conditions (BDic) to the present case in 
which the potential felt by the fluctuations can never be neglected.

In general, the mode functions for large $ k $ behave as 
free modes since the potential $ W_\mathcal{R} $ 
 becomes negligible in this limit except at the singularity
$ t=t_* $. One can then impose
Bunch-Davies conditions for large $ k $ which corresponds to
assume an initial quantum vacuum Fock state, empty of curvature excitations
\be\label{BDkIntro}
S_\mathcal{R}(k;\eta) \buildrel{k \to \infty }\over= 
\frac{e^{-i \; k \; \eta}}{\sqrt{2 \; k}}
\ee
and therefore
$$
\frac{dS_\mathcal{R}}{d \eta}(k;\eta_0) \buildrel{k \to \infty }\over= 
-i \; k \; S_\mathcal{R}(k;\eta_0)\; .
$$
Here $ \eta $ stands for the conformal time: $ d\eta = dt / a(t) $.
Eq.(\ref{BDkIntro}) fulfils the Wronskian normalization 
(that ensures the canonical commutation relations)
\be\label{wronskianIntro}
W[S_\mathcal{R},S^*_\mathcal{R}]= S_\mathcal{R} \; 
\frac{dS^{*}_\mathcal{R}}{d \eta} - 
\frac{dS_\mathcal{R}}{d \eta} \; S^*_\mathcal{R} = i \; .
\ee
In asymptotically flat (or conformally flat) regions of the space-time
the potential felt by the fluctuations $ W_\mathcal{R}(\eta) $ vanishes and the fluctuations
exhibit a plane wave behaviour for {\bf all} $ k $ (not necesarily large).
This is not the case in strong gravity fields or near curvature singularities
as in the present cosmological space-time where $ W_\mathcal{R}(\eta) $ can 
 never be neglected at fixed $ k $. However, we can choose Bunch-Davies initial 
conditions (BDic) at $ \eta = \eta_0 $  (or equivalently, $ t = t_0 $) by imposing
\be\label{BD1Intro}
\frac{dS_\mathcal{R}}{d \eta}(k;\eta_0) = -i \; k \; 
S_\mathcal{R}(k;\eta_0) \quad {\rm for ~ all} \; k \; .
\ee
That is, we consider the initial value problem for the mode functions 
giving the values of $ S_\mathcal{R}(k;\eta) $ and 
$ dS_\mathcal{R}/d \eta $ at $ \eta = \eta_0 $.
This condition combined with the Wronskian condition 
eq.(\ref{wronskianIntro}) implies that
\be\label{BD2Intro}
|S_\mathcal{R}(k;\eta_0)| = \frac1{\sqrt{2 \; k}} \quad , \quad
\left| \frac{dS_\mathcal{R}}{d \eta}(k;\eta_0) \right| = 
\sqrt{\frac{k}2} \; .
\ee
which is equivalent to eq.(\ref{BDkIntro}) for large $ k $.

\medskip 

The power spectrum at the end of slow-roll inflation $ P_\mathcal{R}(k) $
gets dynamically modified by the effect of the preceding fast-roll eras 
through the transfer function of initial conditions $ D(k) $:
\be \label{powRIntro}
P_\mathcal{R}(k)= P^{BD}_{\mathcal{R}}(k)\left[1+ D(k) \right] \; ,
\ee 
 
$ D(k) $ accounts for the effect of both the initial conditions 
and the fluctuations evolution during fast-roll (before slow-roll).
$ D(k) $ depends on the time $ t_0 $ at which BDic are imposed.

\medskip

The power spectrum  $ P^{BD}_{\mathcal{R}}(k) $ corresponds to start the evolution
with pure slow-roll from $ t_0 \to-\infty $  and 
with BDic eq.(\ref{BD1Intro})-eq.(\ref{BD2Intro}) imposed 
there at $ t_0 \to-\infty $, that is $ \eta_0 =-\infty $. 
$ P^{BD}_{\mathcal{R}}(k) $ is given by its customary pure slow-roll expression,
\be\label{pbdIntro}
 \log  P^{BD}_{\mathcal{R}}(k) = \log A_s(k_0) + (n_s-1) \; \log\frac{k}{k_0} + 
\tfrac12 \; n_{run} \;  \log^2\frac{k}{k_0} +{\cal O}\left(\frac1{N^3}\right) \; .
\ee
where $ N $ is the number of inflation efolds since the pivot CMB scale $ k_0 $ exits
the horizon. We take here $ N=60 $.

Actually, BDic can be imposed at $ \eta = \eta_0 = -\infty $ if and only if
the inflaton evolution {\bf also} starts at $ \eta = \eta_0 = -\infty $.
This {\bf only} happens for a {\it particular} inflaton solution: the {\it extreme slow-roll}
solution that we explicitly present and analyze in sec. \ref{esr}.
In the extreme slow--roll case the fast-roll eras are absent,
BDic are imposed at $ t_0 \to-\infty $ (that is $ \eta_0 =-\infty $), then
$ D(k) = 0 $ and $ P_\mathcal{R}(k) = P^{BD}_{\mathcal{R}}(k) $. Only in this case
the fluctuation power spectrum at the end of inflation is the usual power spectrum  
$ P^{BD}_{\mathcal{R}}(k) $ eq.(\ref{pbd}).
 
\medskip 
 
When BDic are imposed at {\bf finite times} $ t_0 $, the spectrum {\bf is not} the
usual $ P^{BD}_{\mathcal{R}}(k) $ but it gets modified by 
a non-zero transfer function $ D(k) $ 
eq.(\ref{powR}). The power spectrum $ P_{\mathcal{R}}(k) $ vanishes at $ k = 0 $  
and exhibits oscillations which vanish at large $ k $ [see figs. \ref{fig:DBDC} and \ref{fig:BDs}]

\medskip

Generically, the power spectrum vanishes at $ k = 0 $ and we thus have 
\be\label{propgIntro}
1 + D(k) \buildrel{k \to 0 }\over= {\cal O}(k^{n_s+1}) \quad .
\ee
as shown in sec. \ref{dcero}.
$ D(k) $ presents a first peak for $ k \sim 2/\eta_0 $ and then oscillates asymptotically 
with decreasing amplitude such that
\be\label{propgIntro2}
D(k) \buildrel{k \to \infty }\over= {\cal O}\left(\frac1{k^2}\right) \; .
\ee

\medskip 

We solved numerically the fluctuations equation 
with the BDic eq.(\ref{BD}) covering both the fast-roll and slow-roll
regimes, namely for different initial times $ t_0 $ ranging from the singularity 
$ \tau = \tau_* $ till the transition time $ \tau_{trans} $
from fast-roll to slow-roll. That is to say, 
we solved the fluctuations evolution for BDic imposed at different times in the three eras 
and we compare the resulting power spectra among them.
We computed the corresponding transfer function, $ D(k) $
for the  BDic imposed at the different eras.
We depict  $ 1 + D(k) $ vs. $ k $  for the different
values of the time $ t_0 $ where BDic are imposed in figs. \ref{fig:DBDC}.

\medskip

When the BDic are imposed during the fast--roll stage well {\bf before} it ends, 
$ D(k) $ changes much more significantly than along the extreme slow roll solution. 
This is due to two main effects: the
potential felt by the fluctuations is attractive during fast--roll, and $ \eta_0 $, (far from 
being almost proportional to $ 1/a(\eta) $), tends to the constant value 
$ \eta_\ast $ as $ \tau \to \tau_\ast^+ $ and $ a(\eta)\to 0 $. The numerical 
transfer functions $ 1+ D(k) $ obtained from eqs.(\ref{curvapot}) and (\ref{powR})
are plotted in figs.~\ref{fig:DBDC}.

\medskip
 
We have also computed $ D(k) $ analytically with BDic at finite times $ \eta_0 $, and a simple
form is obtained in the scale invariant case, 
which is the leading term in the slow-roll expansion:
\be\label{dsr32Intro}
D(k) = \frac{\cos 2x}{x^2} - \frac{\sin 2x}{x^3}  + \frac{\sin^2 x}{x^4}
\quad , \quad   x \equiv k \; \eta_0 \; .
\ee

\medskip

Different initial times $ t_0 $ lead essentially to a rescaling of $ k $
in $ D(k) $ by a factor  $ \eta_0 $ since the 
conformal time $ \eta $ is almost proportional to $ 1/a(\eta) $ during slow-roll 
[see figs.~\ref{fig:DBDC} and below eq.(\ref{dsr})]. 
By virtue of the dynamical attractor character of slow--roll, 
the power spectrum when the BDic are imposed at a finite time 
$ t_0 $ cannot really distinguish between the extreme slow--roll solution or any other 
solution which is attracted to slow--roll well before the time $ t_0 $. 

\medskip

Using the transfer function $ D(k) $ we obtained, we computed the change on the 
CMB multipoles $ \Delta C_{\ell}/ C_{\ell} $ for $ \ell =1, 2 $ and $3$ as
functions of the starting instant of the fluctuations $ t_0 $.
We plot $ \Delta C_{\ell}/ C_{\ell} $ for $ 1 \leq \ell \leq 5 $
vs. $ t_0 - t_\ast $ in fig. \ref{dcl}. We see that $ \Delta C_{\ell}/ C_{\ell} $ is {\bf positive} for 
small $ t_0  - t_\ast $ and {\bf decreases} with $ t_0 $ becoming then 
{\bf negative}. The CMB quadrupole observations indicate a large {\bf suppression}
thus indicating that $ t_0 - t_\ast \gtrsim 0.05/m \simeq 10100 \; t_{Planck}$.

\medskip

The fact that choosing BDic leads to a primordial power and its respective
CMB multipoles which correctly {\bf reproduce} the observed spectrum 
justifies the use of BDic.

\medskip

Besides finding a CMB quadrupole suppression in agreement with observations 
\cite{biblia}-\cite{quamc},
we provide here {\bf predictions} for the dipole and $ \ell \leq 5 $-multipole 
suppressions. Forthcoming CMB observations can provide better data to confront our
CMB multipole suppression predictions.
It will be extremely interesting to measure the primordial dipole and compare
with our predicted value.

\section{The pre-inflationary and inflationary fast-roll eras}\label{uno}

The current WMAP data are validating the single field slow-roll scenario 
\cite{WMAP5}. Single field slow-roll models provide an appealing, 
simple and fairly generic description of inflation. This 
inflationary scenario can be implemented using a scalar field, 
the \emph{inflaton} with a Lagrangian density (see for example ref. \cite{biblia})
\be
\mathcal{L} = a^3(t)\left[\frac{\dot{\varphi}^2}2 -
\frac{(\nabla\varphi)^2}{2 \, a^2(t)}-V(\varphi) \right] \; ,
\ee 
where $ V(\varphi) $ is the inflaton potential. Since the universe
expands exponentially fast during inflation, gradient terms 
are  exponentially suppressed and can be neglected.
At the same time, the exponential stretching of spatial lengths
classicalize the physics and permits a classical treatment.
One can therefore consider an homogeneous and classical inflaton field 
$ \varphi(t) $ which obeys the evolution equation
\be\label{eqno} 
{\ddot \varphi} + 3 \, H(t) \; {\dot \varphi} + V'(\varphi) = 0 
\ee 
in the isotropic and homogeneous FRW metric which is sourced by the inflaton
\be\label{FRW}
 ds^2= dt^2-a^2(t) \; d\vec{x}^2
\ee
$ H(t) \equiv {\dot a}(t)/a(t) $ stands for the Hubble parameter.
The energy density and the pressure for a spatially homogeneous inflaton 
are given by
\be\label{enerpres1} 
\rho = \frac{\dot{\varphi}^2}2+ V(\varphi)
\quad , \quad p  =\frac{\dot{\varphi}^2}2-V(\varphi) \; . 
\ee
Threfore, the scale factor $ a(t) $ obeys the Friedmann equation,
\be\label{frinf}
H^2(t) = \frac1{3 M^2_{Pl}} \left[\frac12 \; \dot \varphi^2 + 
V(\varphi)\right] \; .
\ee
In order to have a finite number of inflation efolds, the inflaton 
potential $ V(\varphi) $ must vanish at its absolute minimum
\be\label{minV}
V'(\varphi_{min})=V(\varphi_{min}) = 0
\ee
Otherwise, inflation continues forever.

\medskip 

We formulate inflation as an effective field theory within the 
Ginsburg-Landau spirit \cite{1sN,gl,biblia}. The theory of the second order 
phase transitions, the Ginsburg-Landau theory of superconductivity, 
the current-current Fermi theory of weak interactions, the sigma model of 
pions, nucleons (as skyrmions) and photons are all successful
effective field theories. Our work shows how powerful is
the effective theory of inflation {\bf to predict observable quantities} 
that can be or will be soon contrasted with experiments. 

\medskip 

The effective theory of inflation should be the 
low energy limit of a microscopic fundamental
theory not yet precisely known. The energy scale of inflation $ M $ should be
at the Grand Unified Theory (GUT) energy scale in order to 
reproduce the amplitude of the CMB anisotropies \cite{biblia}. 
Therefore, the microscopic theory of inflation 
is expected to be a GUT in a cosmological space-time.
Such a theory of inflation would contain many fields of various spins.
However, in order to have a 
homogeneous and isotropic universe the expectation value of the 
energy-momentum tensor of the fields must be homogeneous and isotropic. 
The inflaton field in the effective theory may be a coarse-grained average of
fundamental scalar fields, or a composite (bound state) of fundamental fields
of higher spin, just as in superconductivity. The inflaton does
not need to be a fundamental field, for example it may emerge as
a condensate of fermion-antifermion pairs $ < {\bar \Psi} \Psi> $
in a GUT in the cosmological background. In
order to describe the cosmological evolution is enough to consider
the effective dynamics of such condensates.  The relation between
the effective field theory of inflation and the
microscopic fundamental GUT is akin to the relation between the
effective Ginzburg-Landau theory of superconductivity and the
microscopic BCS theory, or like the relation of the $O(4)$ sigma
model, an effective low energy theory of pions, photons and chiral
condensates with quantum chromodynamics (QCD) \cite{quir}.

Vector fields have been considered to describe inflation in
ref.\cite{gmv}. The results for the inflaton should not be very different from the 
effective inflaton description since the energy-momentum tensor
of the vector field is to be taken homogeneous and isotropic.
Namely, we are always in the presence of a scalar condensate.

Since the mass of the inflaton is given by $ M^2/M_{Pl} \sim 10^{13} $GeV 
\cite{biblia}, massless fields alone cannot describe inflation which 
leads to the observed amplitude of the CMB anisotropies.

\medskip 

The classical inflaton potential $ V(\varphi) $ gets modified by
quantum loop corrections. 
We computed relevant quantum loop corrections to inflationary dynamics
in ref. \cite{biblia,effpot}.  A
thorough study of the effect of quantum fluctuations reveals that
these loop corrections are suppressed by powers of $
\left(H/M_{Pl}\right)^2 \sim 10^{-9} $ where $ H $ is the Hubble parameter during 
inflation \cite{biblia,effpot}.  Therefore, quantum loop corrections 
are very small, a conclusion that validates the reliability of the classical
approximation and of the effective field theory approach to
inflationary dynamics. In particular, the (small) one-loop 
corrections to the potential in an inflationary universe are very 
different from the Coleman-Weinberg form \cite{biblia,effpot}. 

\medskip 

We choose the inflaton field initially homogeneous which
ensures it is always homogeneous. The fluctuations around
are small and give small corrections to the homogeneity
of the Universe. The rapid expansion of the Universe,
in the inflationary regimes, takes care
of the classical fluctuations, quickly flattening an eventually
non-homogeneous condensate.

\subsection{The complete inflaton evolution through the different eras}

It is convenient to use the dimensionless variables to analyze
the inflaton evolution equations eqs.(\ref{eqno})-(\ref{frinf}),
\cite{biblia}:
\be \label{tau} 
\tau =  m \; t \quad , \quad h \equiv \frac{H}{m}  
\quad , \quad \phi = \frac{\varphi}{M_{Pl}} \; .
\ee 
The inflaton potential has then the universal form
\be\label{v1}
V(\varphi) = M^4 \; v\left(\frac{\varphi}{M_{Pl}}\right)  \; ,
\ee
where $ M $ is the energy scale of inflation and $ v(\phi) $ is a dimensionless function.
Without loss of generality we can set $ v'(0)=0 $ \cite{biblia}. Moreover, provided 
$ V''(0)\neq 0 $ we can set without loss of generality $ |v''(0)|=1/2 $.
Namely, we have for small fields,
\be\label{vpol}
v(\phi) \buildrel{\phi \to 0 }\over= v(0) \mp \frac12 \; \phi^2 + 
{\cal O}(\phi^3)
\ee
where the minus sign in the quadratic term corresponds to new inflation
and the plus sign to chaotic inflation.

In these dimensionless variables, the energy density and the pressure 
for a spatially homogeneous inflaton are given from eq.(\ref{enerpres1}) by
\be\label{enerpres} 
\frac{\rho}{M^4} = \frac12 \left(\frac{d\phi}{d \tau}\right)^2 + v(\phi)
\quad , \quad \frac{p}{M^4}  =  \frac12 \left(\frac{d\phi}{d \tau}\right)^2
-v(\phi) \; ,
\ee
and the coupled inflaton evolution equation (\ref{eqno})
and the Friedmann equation (\ref{frinf}) take the form \cite{biblia},
\bea \label{evol} 
&& \frac{d^2 \phi}{d \tau^2} + 3 \; h \; \frac{d\phi}{d \tau} + 
v'(\phi) = 0 \quad , \cr \cr
&&  h^2(\tau) = \frac13\left[\frac12 \left(\frac{d\phi}{d \tau}\right)^2 
+ v(\phi) \right] \quad .
\eea 
These coupled nonlinear differential equations completely define the time
evolution of the inflaton field and the scale factor once the initial
conditions are given at the initial time $ \tau_0 $. Namely, the initial
conditions are fixed by giving two real 
numbers, the values of $ \phi(\tau_0) $ and $ d\phi(\tau_0)/d\tau $.

\medskip

It follows from eqs.(\ref{evol}) that 
\be\label{aseg}
\frac{d^2 a}{d \tau^2} = \frac13\left[
v(\phi)-\left(\frac{d\phi}{d \tau}\right)^2\right] = -\frac12 \left(p
+\frac13 \; \rho \right) \; .
\ee
When $ d^2 a/d \tau^2 > 0 $ the expansion of the universe accelerates
and it is then called inflationary. 

\medskip

The derivative of the Hubble parameter is {\bf always} negative:
\be\label{hpu}
\frac{dh}{d \tau} = -\frac12 \left(\frac{d\phi}{d \tau}\right)^2 \; .
\ee
Therefore $ h(\tau) $ decreases monotonically with increasing $ \tau $.
Conversely, if we evolve the solution backwards in time from $ \tau_0, \;
 h(\tau) $ will generically {\bf increase} without bounds. Namely, at some time 
$ \tau = \tau_* , \;  h(\tau) $ can exhibit a singularity
 where simultaneously $ a(\tau_*) $ vanishes.

\medskip

In fact, the equations (\ref{evol}) admit the singular solution for 
$ \tau \to \tau_* $,
\be\label{sing1}
\phi(\tau)\buildrel{\tau \to \tau_* }\over= \sqrt{\frac23} \; 
\log\frac{\tau-\tau_*}{b} \to -\infty \quad , \quad 
h(\tau) \equiv \frac{d}{d \tau} \log a(\tau)
\buildrel{\tau \to \tau_* }\over= \frac1{3 \; (\tau-\tau_*)}  
\to +\infty \; ,
\ee
where $ b $ is an integration constant. The energy density $ \epsilon(\tau) $ and equation of state
take the limiting form,
\be\label{pe2}
\rho(\tau)\buildrel{\tau \to \tau_* }\over=\frac1{3 \; (\tau-\tau_*)^2}
\to +\infty 
\quad , \quad \frac{p(\tau)}{\rho(\tau)}\buildrel{\tau \to \tau_* }\over=1
\; .
\ee
Namely, the limiting equation of state is $ p \buildrel{\tau \to \tau_* }\over= + \rho $.

We have in this regime
\be\label{acero}
a(\tau) \buildrel{\tau \to \tau_* }\over= C \; (\tau-\tau_*)^\frac13 
\to 0 \; ,
\ee
where $ C $ is some constant.
That is, the geometry becomes singular for $ \tau \to \tau_* $.
The behaviour near $ \tau_* $ is non-inflationary, namely decelerated, since
\be\label{asegs}
 \frac{d^2 a}{d \tau^2}\buildrel{\tau \to \tau_* }\over= -\frac29 \;
 (\tau-\tau_*)^{-\frac53} \to -\infty \; .
\ee
For  $ \tau \to \tau_* $, near the singularity,
the potential $ v(\phi) $ becomes negligible 
in eqs.(\ref{evol}). Therefore,  eqs.(\ref{sing1})-(\ref{asegs}) are valid for
all regular potentials $ v(\phi) $.

\medskip

\begin{figure}[h]
\includegraphics[width=16.cm]{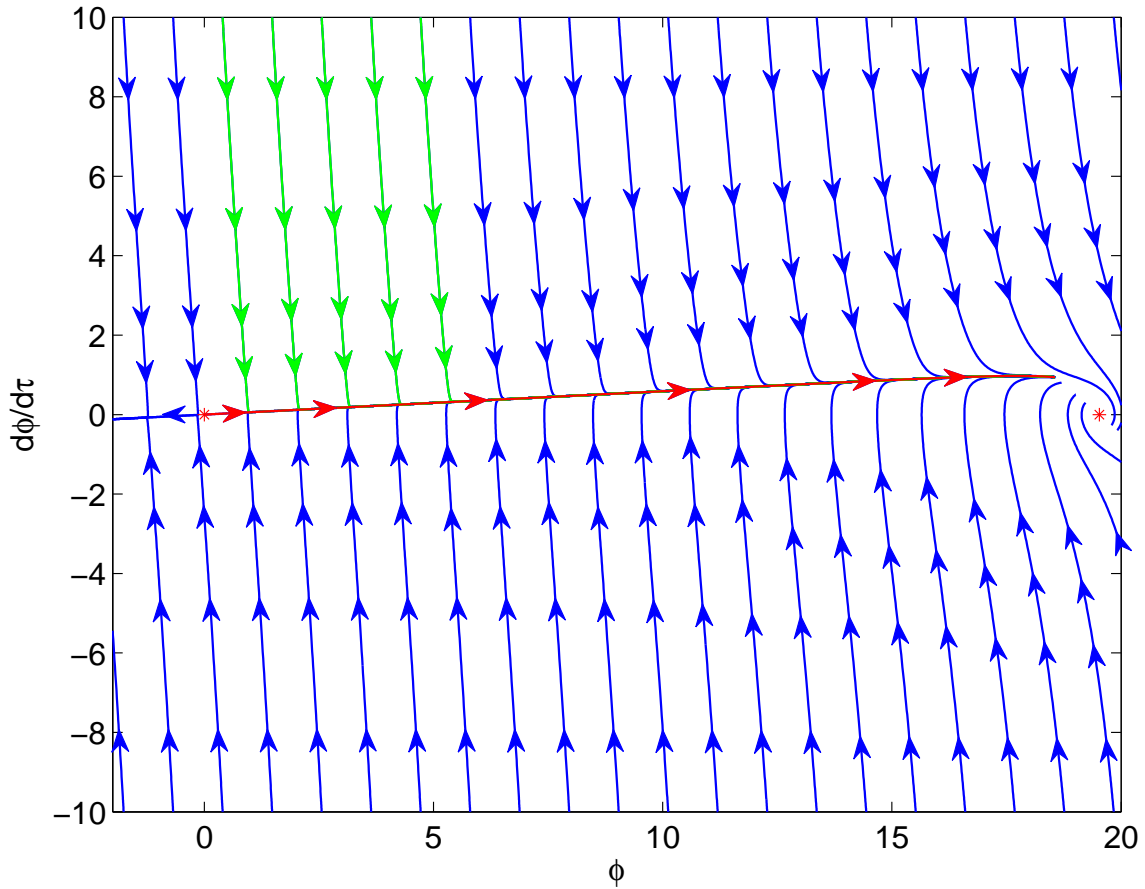}
\caption{The complete inflaton flow in phase space. 
$ d \phi/d \tau $ vs. $ \phi $ for different initial conditions.
We see that the inflaton always reaches a slow-roll regime for generic initial
conditions represented by a red quasi-horizontal line. Hence,
the slow-roll line is an attractor.
Ultimately the inflaton reaches asymptotically the absolute minima $ d\phi/d \tau = 0 ,
\; \phi = \phi_{min} =  \sqrt{8 \, N/y} =  19.52\ldots$. The number of efolds of slow-roll 
inflation $ N_{sr} $ increases for decreasing initial $ \phi> 0 $ when $ d\phi/d \tau > 0 $ 
initially. The $ \phi> 0 , \; d\phi/d \tau > 0 $ trajectories corresponding to $ N_{sr} > 63 $ 
are colored in green.} 
\label{phs}
\end{figure}

The evolution starts thus by 
this decelerated fast-roll regime followed by an inflationary fast-roll
regime and then by a slow-roll inflationary regime \cite{biblia}.
Recall that the slow-roll regime is an {\bf attractor} \cite{bgzk}, and therefore
the inflaton always reaches a slow-roll inflationary regime for
{\bf generic} initial conditions. We display in fig. \ref{phs} the inflaton flow
in phase space, namely $ d \phi/d \tau $ vs. $ \phi $ for different initial conditions.

The number of efolds of slow-roll inflation $ N_{sr} $ is determined by the
time when the inflaton trajectory reaches the red quasi-horizontal line of slow-roll regime
[see fig. \ref{phs}]. 
We see that $ d \phi/d \tau $ decreases steeply with $ \phi $. This implies
that $ N_{sr} $ is mainly determined by the initial value of $ \phi $
with a mild (logarithmic) dependence on the initial value of $ d \phi/d \tau $

The inflaton flow described by eq.(\ref{sing1}) results
\be\label{fifis}
{\dot \phi}(\tau) \buildrel{\tau \to \tau_* }\over=\sqrt{\frac23} \; 
\frac{e^{-\sqrt{\frac32} \; \phi(\tau)}}{b}
\ee
which well reproduce the almost vertical blue and green lines in fig. \ref{phs}.

\medskip

The inflationary regimes are characterized by 
the slow-roll parameters $ \epsilon_v $ and $ \eta_v $ \cite{biblia}
\be\label{epsv}
\epsilon_v =  \frac1{2 \; h^2} \; 
\left(\frac{d \phi}{d \tau}\right)^2 
\quad , \quad \eta_v = \frac{v''(\phi)}{v(\phi)} \; .
\ee
The slow-roll behaviour is defined by the condition $ \epsilon_v < 1/N $.
Typically, $ \epsilon_v \lesssim 1/N $ during slow-roll. More generally
accelerated expansion (inflation) happens for $ \epsilon_v < 1 $ while
we have decelerated expansion for $ \epsilon_v > 1 $ as follows
from eqs.(\ref{enerpres})-(\ref{aseg}) and (\ref{epsv}).

The parameter $ \eta_v $ is also of the order $ 1/N $ during slow-roll and
it is generically of order $ 1/N $ during fast-roll except when the potential $ v(\phi) $
vanishes.

\medskip

Eq.(\ref{hpu}) implies a monotonic decreasing of the expansion rate of the
universe. There are {\bf four stages} in the universe evolution
described by eqs.(\ref{evol}):

\begin{itemize}
\item{The non-inflationary fast-roll stage starting at the singularity
$ \tau = \tau_* $ and ending when  $ d^2 a/d \tau^2 $ becomes positive
[see eq.(\ref{aseg})].}
\item{The inflationary fast-roll stage starts when $ d^2 a/d \tau^2 $ 
becomes positive and ends at $ \tau = \tau_{trans} $ when  $ \epsilon_v $ 
becomes smaller than $ 1/N $ [see eq.(\ref{epsv})].}
\item{The inflationary slow-roll stage follows, and it continues as long as
$ \epsilon_v < 1/N $ and $ d^2 a/d \tau^2 > 0 $. It ends when 
$ d^2 a/d \tau^2 $ becomes negative at $ \tau = \tau_{end} $.}
\item{A matter-dominated stage follows the inflationary era.}
\end{itemize}

The four stages described above correspond to the evolution for generic
initial conditions or, equivalently, starting from the singular
behaviour eqs.(\ref{sing1}). In addition, there exists a special (extreme) 
slow-roll solution starting at $ \tau = -\infty $ where the fast-roll stages are 
absent. We derive this extreme slow-roll solution in sec. \ref{esr}.

\medskip

As shown in refs. \cite{mcmc,biblia} the double well (broken symmetric) 
fourth order potential 
\be\label{bini}
V(\varphi) = \frac14 \; \lambda \; 
\left(\varphi^2-\frac{m^2}{\lambda}\right)^2=
- \frac12 \, m^2 \; \varphi^2 + \frac14 \; \lambda \;\varphi^4 + 
\frac{m^4}{4 \, \lambda}
\ee
provides a very good fit for the CMB+LSS data, while at the same time being 
particularly simple, natural and stable in the Ginsburg-Landau sense.
This is a new inflation model with the inflaton rolling from the vicinity
of the local maxima of $ V(\varphi) $ at $ \varphi = 0 $ towards the
absolute minimum $ \varphi = m/\sqrt{\lambda} $.

The inflaton mass $ m $ and coupling $ \lambda $ 
are naturally expressed in terms of the 
{\bf two} relevant energy scales in this problem: the energy scale of inflation $ M $
and the Planck mass $ M_{Pl} = 2.43534 \; 10^{18}$ GeV,
\be
m=\frac{M^2}{M_{Pl}} \;,\quad \lambda = \frac{y}{8 \, N} \; 
\left(\frac{M}{M_{Pl}}\right)^4 \; .
\ee
Here $ N \sim 60 $ is the 
number of efolds since the cosmologically relevant modes exit the horizon 
till the end of inflation and $ y \sim 1 $ is the quartic coupling.

The MCMC analysis of the CMB+LSS
data combined with the theoretical input above yields the value 
$ y \simeq 1.26 $ for the coupling \cite{mcmc,biblia}.
$ y $ turns to be {\bf order one} consistent with the Ginsburg-Landau 
formulation of the theory of inflation \cite{biblia}.

This model of new inflation yields as most probable 
values: $ n_s \simeq 0.964 ,\; r\simeq 0.051 $ \cite{mcmc,biblia}. 
This value for $ r $ is within reach of forthcoming CMB observations. 
For $ y > 0.431946\ldots $ 
and in particular for the best fit value $ y \simeq 1.26 $,
the inflaton field exits the horizon in the negative
concavity region $ V''(\varphi) < 0 $ intrinsic to new inflation \cite{biblia}.
We find for the best fit \cite{mcmc,biblia}, 
\be\label{masas}
M = 0.543 \times 10^{16} \quad {\rm GeV ~ for ~ the ~ scale ~ of  ~ 
inflation ~ and} \quad m = 1.21 \times 10^{13} \quad {\rm GeV ~ for ~ 
the ~ inflaton ~ mass.}
\ee

\medskip

We consider from now on the quartic broken symmetric potential eq. (\ref{bini})
which becomes using eq.(\ref{v1})] 
\be\label{vnue}
v(\phi) = \frac{g}4 \left(\phi^2 - \frac1{g}\right)^{\! 2} 
= -\frac12 \; \phi^2 + \frac{g}4 \; \phi^4 + \frac1{4 \; g} \quad {\rm where} \quad
g = \frac{y}{8 \; N} \; .
\ee
We have two arbitrary real coefficients characterizing the initial 
conditions. We can choose them as $ b $ and $ \tau_* $ [see  eq.(\ref{sing1})]. 
A total number of slow-roll inflation efolds $ N_{sr} \simeq 63 $ permits to
explain the CMB quadrupole suppression \cite{quadru,quamc,biblia}. Such 
requirement fixes the value of  $ b $ for a given coupling $ y $.

We integrated numerically eqs.(\ref{evol}) with eq.(\ref{sing1}) as 
initial conditions. 
We find that $ b = 4.745272\ldots \; 10^{-5} $ yields 63 efolds of inflation 
during the slow-roll era for $ y = 1.26 $,  the best fit to
the CMB and LSS data.
We find that $ b $ is a monotonically increasing function of the coupling 
$ y $ for fixed number of slow-roll efolds. At fixed coupling, $ b $
increases with the number of slow-roll efolds. 

\begin{figure}[h]
\includegraphics[width=16.cm]{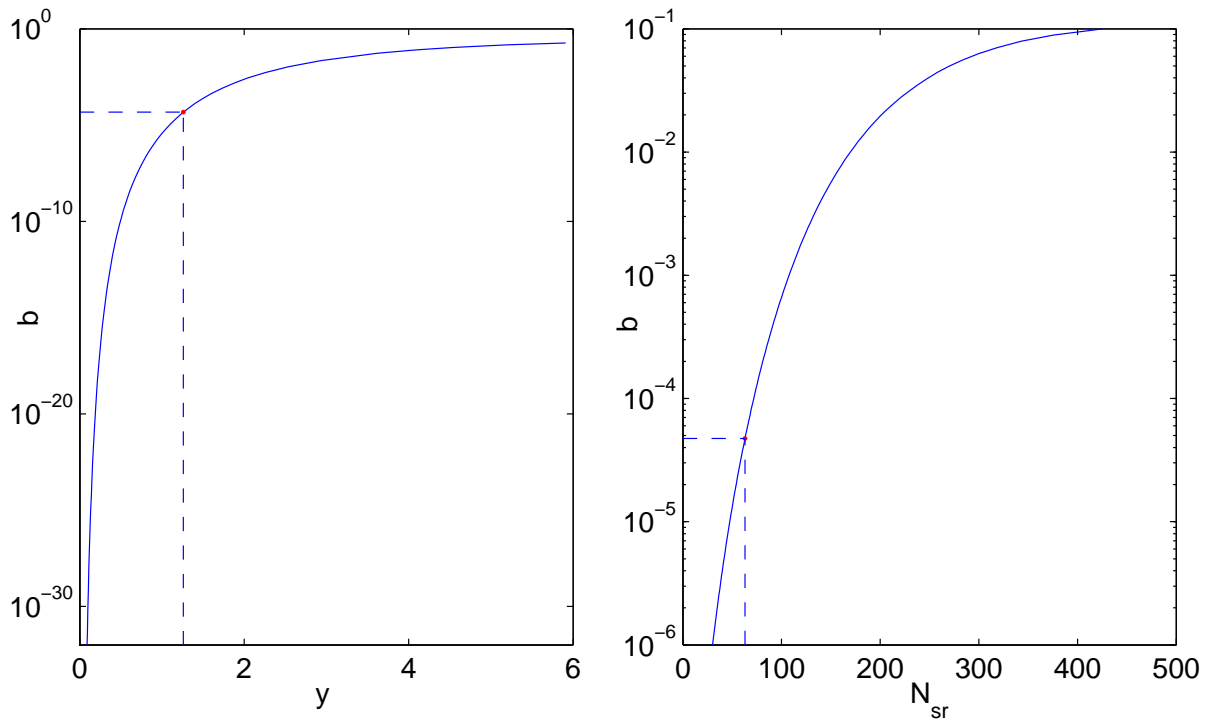}
\caption{Left panel: the coefficient $ b $ characterizing the initial conditions
  vs. the quartic coupling $ y $ for $ N_{sr} = 63 $ efolds of slow-roll
  inflation. Right panel: b vs. $N_{sr}$ for $y=1.26$. The preferred values
  $y=1.26$ and $N_{sr}$ are highlighted in both panels.}
\label{ylnb}
\end{figure}

We display in fig. \ref{ylnb} $ b $ as a function of $ y $ and the number of
slow--roll inflation efolds $ N_{sr} $. 

\begin{figure}[h]
\includegraphics[width=16.cm]{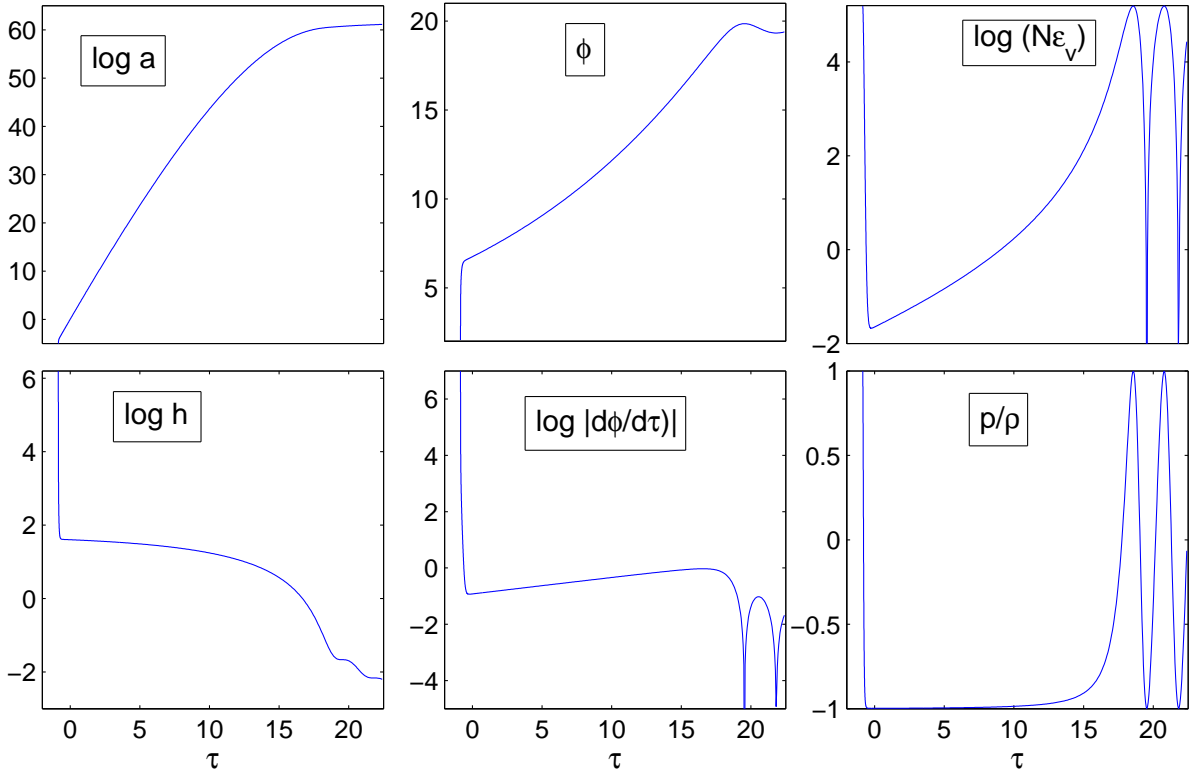}
\caption{Time evolution during during the three eras: non-inflationary fast-roll, 
inflationary fast-roll and slow-roll and beyond the end of inflation (MD era). 
$ \log a(\tau) , \;  \log h(\tau), \; \phi(\tau), \; \log|{\dot \phi}(\tau)|, \;
 \log [N \; \epsilon_v(\tau)] $ and $ p(\tau)/\rho(\tau) $ vs. $ \tau $.
$ a(\tau) $ grows monotonically reaching 63 efolds by the end of inflation.
$ h(\tau) $ diverges for $ \tau \to \tau_* = -0.8499574\ldots $ according to
eq.(\ref{sing1}) and decreases fast during fast-roll
($ \tau \leq \tau_{trans} = 0.2487963\ldots $). Then, $ h(\tau) $ 
decreases slowly during slow-roll as discussed in sec. \ref{solusr}.
We depict $ h(\tau) $ for short times ($ 0 < \tau- \tau_* < 0.3 $) in fig. \ref{apefi}.
$ {\dot \phi}(\tau) $ diverges for $ \tau \to \tau_* $ according to
eq.(\ref{sing1}) and decreases fast during fast-roll becoming
very small during slow-roll. After the fast-roll stage where 
the inflaton field grows according to eq.(\ref{sing1}), $ \phi(\tau) $
slowly rolls toward its absolute minimum at $ \phi_{end} = \sqrt{8 \, N/y} = 
19.52\ldots $.
$ \log [N \; \epsilon_v(\tau)] $ vs. $ \tau- \tau_* $.
We have that $ \epsilon_v(\tau_*) = 3 $ according to
eqs.(\ref{sing1}) and (\ref{epsv}). $ \epsilon_v(\tau) $
decreases fast during fast-roll becoming of the order $ 1/N $. We {\bf define} 
the end of fast-roll (and beginning of slow-roll) by the condition 
$ N \; \epsilon_v(\tau) \equiv 1 $ which gives $ \tau_{trans} - \tau_* = 0.2487963\ldots $. 
The equation of state $ p(\tau)/\rho(\tau) $ fastly decreases during fast-roll from
the value $ p/\rho = +1 $ for $ \tau \to \tau_* $ [see eq.(\ref{pe2})] 
passing through $ p/\rho =  -1/3 $ at the beginning of fast-roll
inflation [see eq.(\ref{aseg})],
$ \tau = \tau_s = \tau_* + 0.0573 $, and reaching $ p/\rho = -1 $ by the
beginning of slow-roll. $ p/\rho $ vanishes again near the end of slow-roll inflation
by $ \tau_{end} = \tau_* + 18.698\ldots $.} 
\label{evolu}
\end{figure}

For this value of $ y $ and 63 efolds of inflation 
during the slow-roll, fast-roll ends by $ \tau = \tau_{trans} = 0.2487963\ldots $.
In figures \ref{evolu}, we depict 
$ \log a(\tau), \; \log h(\tau), \;
\phi(\tau), \; \log|{\dot \phi}(\tau)|, \; \log[ N \; \epsilon_v(\tau)] $
and $ p(\tau)/\rho(\tau) $ vs. $ \tau $ till a short time after the end of inflation.
We define the time $ \tau_{end} $ when inflation 
ends by the condition $ {\ddot a}(\tau_{end}) = 0 $ which 
gives $ (\tau_{end}- \tau_*) = 18.2547816\ldots $.

\bigskip

Furthermore, we study in this paper the curvature and tensor fluctuations
during the {\bf whole} inflaton evolution in its three succesive
regimes: non-inflationary fast-roll,  
inflationary fast-roll and inflationary slow-roll.

\medskip

The equation for the scalar curvature fluctuations take in conformal time 
$ \eta $ and dimensionless variables the form \cite{biblia}
\be\label{fluces}
\left[\frac{d^2}{d\eta^2}+k^2- 
W_\mathcal{R}(\eta)\right]S_\mathcal{R}(k;\eta) =0 \; . 
\ee 
where $ d\eta = d\tau / a(\tau) $, 
\be\label{defwz}
W_\mathcal{R}(\eta) \equiv \frac1{z} \; \frac{d^2 z}{d \eta^2}
\quad {\rm and}  \quad 
z(\eta) \equiv \frac{a(\eta)}{h(\eta)} \; \frac{d\phi}{d \tau} \; .
\ee
In cosmic time $ \tau $, 
eq.(\ref{fluces}) takes the form
\be\label{flutau}
\left[\frac{d^2}{d\tau^2} +  h(\tau) \; \frac{d}{d\tau} +
\frac{k^2}{a^2(\tau)} - V_\mathcal{R}(\tau)\right]S_\mathcal{R}(k;\tau) 
=0 \; . 
\ee
where 
\bea\label{wtau}
&& V_\mathcal{R}(\tau) \equiv 
\frac{W_\mathcal{R}(\tau)}{a^2(\tau)} = h^2(\tau) \; 
\left[ 2 - 7 \; \epsilon_v + 2 \; \epsilon_v^2 - 
\sqrt{8 \; \epsilon_v} \; \frac{v'(\phi)}{h^2(\tau)}  - 
\eta_v \; (3 -  \epsilon_v)\right] = \cr \cr
&& = h^2(\tau) \; \left[ 2 - 7 \; \epsilon_v + 2 \; \epsilon_v^2\right]
- 2 \; \frac{d \phi}{d \tau} \; \frac{v'(\phi)}{h(\tau)} - v''(\phi) 
\quad ,
\eea
and $ \epsilon_v $ and $ \eta_v $ are given by eq.(\ref{epsv}).

\medskip

\begin{figure}[h]
\includegraphics[width=16.cm]{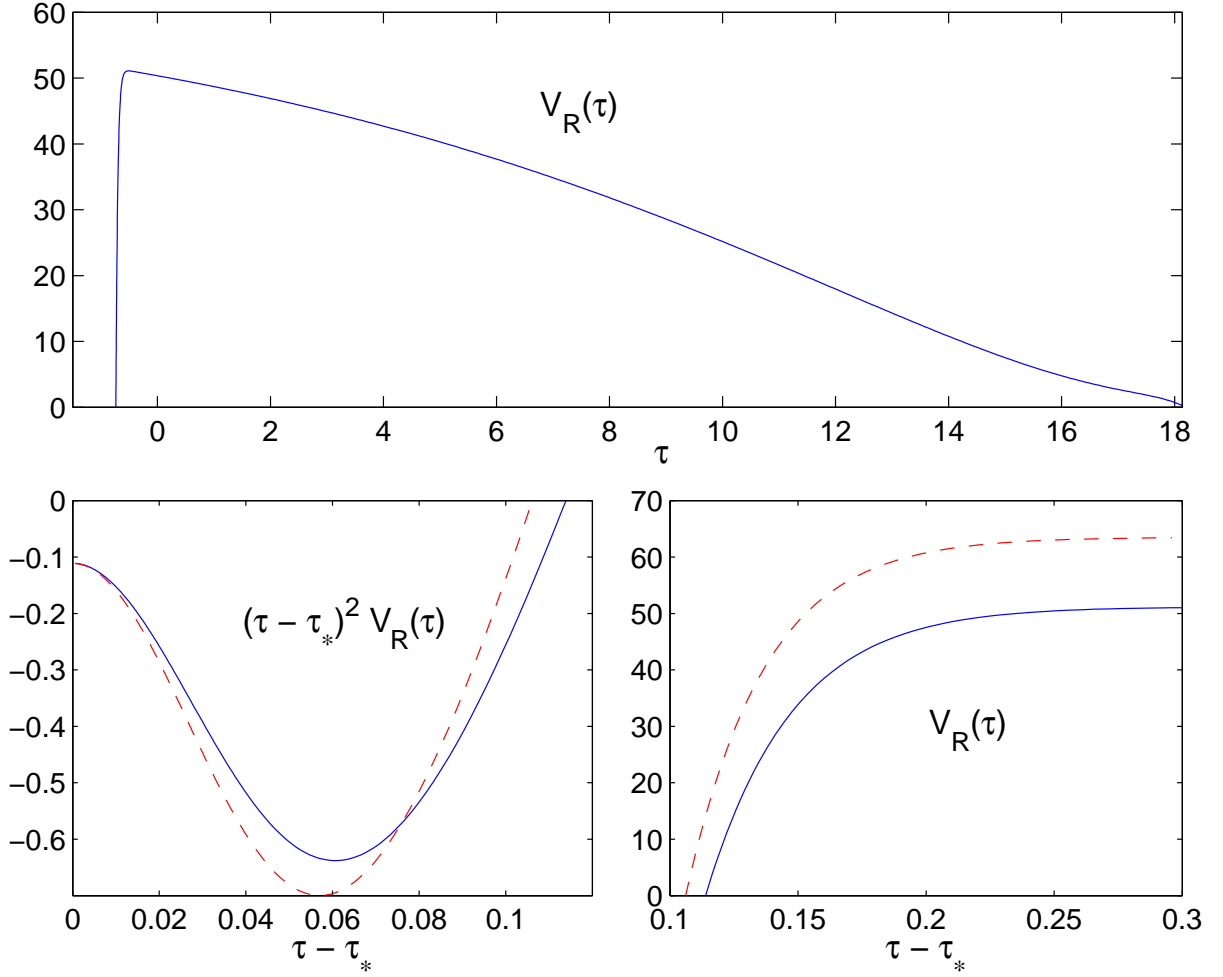}
\caption{The potential $ V_\mathcal{R}(\tau) $ felt by the fluctuations.
Upper plot: $ V_\mathcal{R}(\tau) $  vs. $ (\tau- \tau_*) $ 
in the stage where  $ V_\mathcal{R}(\tau) $ is {\bf repulsive} ($ V_\mathcal{R}(\tau) > 0 $)
which happens for $ (\tau- \tau_*) > 0.114 $. Notice that  $ V_\mathcal{R}(\tau) $
slowly decreases during the slow-roll stage as $  V_\mathcal{R}(\tau) \simeq 2 \; 
h^2(\tau) + 1 + O(1/N) $ according to eq.(\ref{wtau}) and fig.  \ref{evolu}.
Lower plots: Comparison of the exact (numerical) evolution and the analytic approximations
eq.(\ref{vafr}) during fast-roll and slow-roll. Left lower plot: 
$ (\tau- \tau_*)^2 \; V_\mathcal{R}(\tau) $ vs. $ \tau- \tau_* $ 
in the stage where $ V_\mathcal{R}(\tau) $ is
{\bf atractive} ($ V_\mathcal{R}(\tau) < 0 $) from the exact (numerical) calculation
and from the analytic approximation eq.(\ref{vafr}). This happens for 
$ 0 \leq (\tau- \tau_*) < 0.114 $. Notice that  $ \displaystyle 
\lim_{\tau\to \tau_*} (\tau- \tau_*)^2 \; V_\mathcal{R}(\tau) = -1/9 $ according
to eq.(\ref{wep0}). Lower right plot:  $ V_\mathcal{R}(\tau) $  vs. $ \tau- \tau_* $ when 
$ V_\mathcal{R}(\tau) > 0 $ from the exact (numerical) calculation
and from the analytic approximation eq.(\ref{vafr}).} 
\label{potas}
\end{figure}

We display $ V_\mathcal{R}(\tau) $ vs. $ \tau $ in fig. \ref{potas}
for the best fit value of the coupling $ y = 1.26 $ and 63 efolds of slow-roll inflation.

\bigskip

The equation for the tensor fluctuations take in conformal time 
$ \eta $ and dimensionless variables the form \cite{biblia}
\be\label{Sten} 
S^{''}_{T}(k;\eta)+\left[k^2-
\frac{a''(\eta)}{a(\eta)}\right]S_{T}(k;\eta) = 0 \; . 
\ee

\begin{figure}[h]
\includegraphics[width=16.cm]{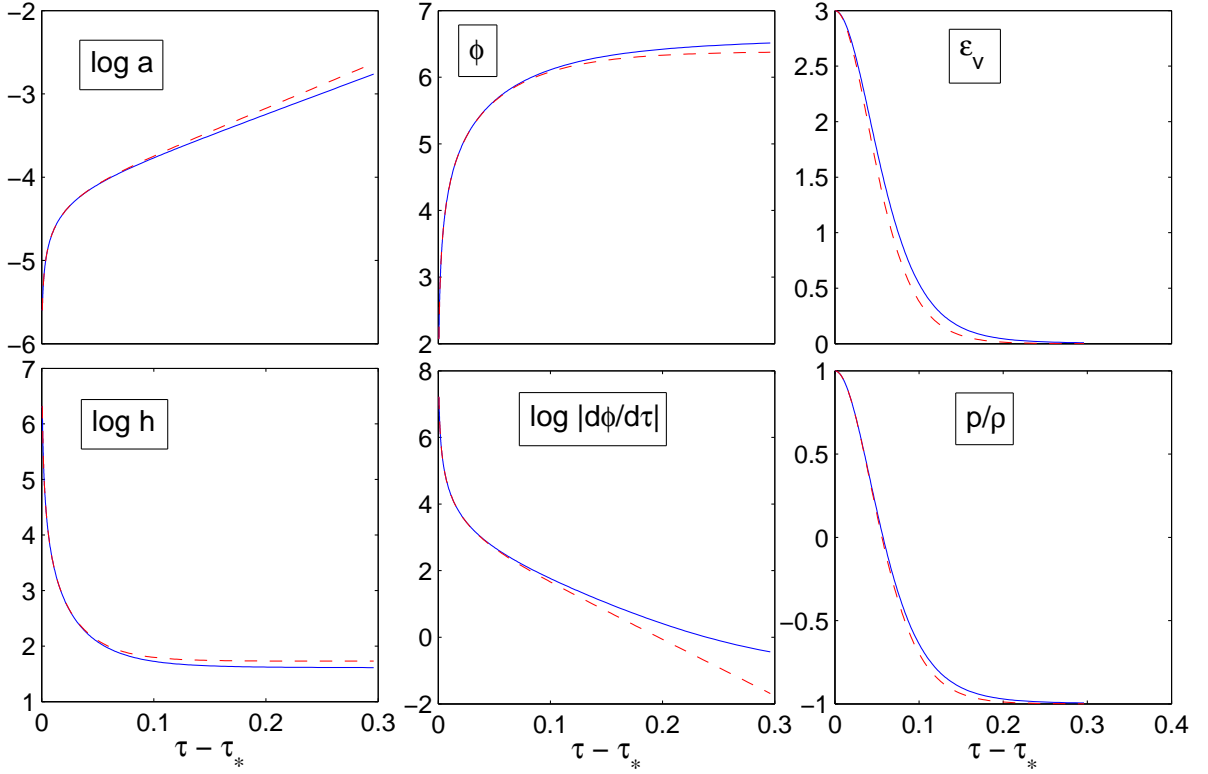}
\caption{Comparison of the exact (numerical) evolution (blue continuous line)
and the analytic approximations (red dashed line) 
eq.(\ref{hafr}) during fast-roll and slow-roll for
$ \log a(\tau) , \;  \log h(\tau), \; \phi(\tau), \; \log|{\dot \phi}(\tau)|, \;
 \epsilon_v(\tau) $ and $ p(\tau)/\rho(\tau) $ vs. $ \tau - \tau_* $. The
exact $ \ln a(\tau) $ and $ \ln h(\tau) $ are close to the approximation eq.(\ref{hafr}). 
The scale factor is normalized to unit at $ \tau = 0 $, sixty efolds before the end of inflation.
The exact (numerical) equation of state $ p(\tau)/\rho(\tau) $ is quite close to
the analytic approximation eq.(\ref{epsfr}) both during fast-roll and slow-roll.
The same happens for the exact (numerical) inflaton field $ \phi(\tau), \; {\dot \phi}(\tau) $
and the analytic approximation eq.(\ref{solfr}).} 
\label{apefi}
\end{figure}

\subsection{Inflaton and scale factor behaviour near the initial mathematical
singularity}\label{persi}

In order to find the behaviour of $ \phi(\tau) $ and $ a(\tau) $
near the initial singularity we write
\be\label{sing2}
\phi(\tau) = \sqrt{\frac23} \; \log\frac{\tau-\tau_*}{b} + \phi_1(\tau)
\quad , \quad 
h(\tau) = \frac1{3 \; (\tau-\tau_*)} + h_1(\tau) \; .
\ee
Inserting now  eqs.(\ref{sing2}) into  eqs.(\ref{sing1}) yields
for $ \phi_1(\tau) $ and $ h_1(\tau) $ the non-autonomous
differential equations
\bea\label{h1fi1}
&& {\ddot \phi}_1 +\left(\frac1{\tau-\tau_*} + 3 \;  h_1 \right){\dot \phi}_1
+ \frac{\sqrt6}{\tau-\tau_*} \; h_1 -  \phi_1 - 
\sqrt{\frac23} \; \log\frac{\tau-\tau_*}{b} + g 
\left(\sqrt{\frac23} \; \log\frac{\tau-\tau_*}{b} + \phi_1\right)^3 = 0 \\ \cr
&& h_1^2 + \frac2{3\; (\tau-\tau_*)}  \; h_1 - \frac{{\dot \phi}_1}6 \; 
\left(\sqrt{\frac23} \; \frac2{\tau-\tau_*} +{\dot \phi}_1 \right)+\frac16 \;
\left(\sqrt{\frac23} \; \log\frac{\tau-\tau_*}{b} + \phi_1\right)^2
-\frac{g}{12} \left(\sqrt{\frac23} \; \log\frac{\tau-\tau_*}{b} + 
\phi_1\right)^4 - \frac1{12 \; g} = 0 \; , \nonumber
\eea
where $ \dot \phi $ stands for $ d\phi/d\tau $.

The asymptotic solution of eqs.(\ref{h1fi1}) for $ \tau \to \tau_* $
turns to have the dominant form
\be\label{defP}
\phi_1(\tau) \buildrel{\tau \to \tau_* }\over=(\tau-\tau_*)^2 \; 
P_4^{\phi}\left(\log\frac{\tau-\tau_*}{b}\right)
\quad , \quad 
h_1(\tau) \buildrel{\tau \to \tau_* }\over= (\tau-\tau_*) \; 
P_4^{h}\left(\log\frac{\tau-\tau_*}{b}\right)
\ee
where $ P_4^{\phi}(z) $ and $ P_4^{h}(z) $ are fourth degree polynomials
in their arguments. The polynomials turn to be of fourth degree
because the inflaton potential is of fourth degree.
Their explicit expressions follow after calculation
\bea\label{expli}
&&\phi_1(\tau) \buildrel{\tau \to \tau_* }\over=-\frac{(\tau-\tau_*)^2}{\sqrt6} \; 
\left[\frac{g}{18}  \left(\log^4\frac{\tau-\tau_*}{b} + 
\frac23 \; \log^3\frac{\tau-\tau_*}{b}
-\frac{11}3 \; \log^2\frac{\tau-\tau_*}{b}+ \frac{49}9 \; \log\frac{\tau-\tau_*}{b}- 
\frac{439}{54}\right) \right. \cr \cr
&& \left. - \frac16 \; \left(\log^2\frac{\tau-\tau_*}{b} + \frac13 \; 
\log\frac{\tau-\tau_*}{b} - \frac78 \right)+ \frac1{8 \; g} \right] \; , \cr \cr
&& h_1(\tau) \buildrel{\tau \to \tau_* }\over=\frac{\tau-\tau_*}9
\left[\frac{g}{18}  \left(6 \; \log^4\frac{\tau-\tau_*}{b} - 
8 \; \log^3\frac{\tau-\tau_*}{b}
+ 8 \; \log^2\frac{\tau-\tau_*}{b} - \frac{11}3 \; \log\frac{\tau-\tau_*}{b} +
\frac{146}9  \right)  \right. \cr \cr
&& \left. - \log^2\frac{\tau-\tau_*}{b}+ \frac23 \; 
\log\frac{\tau-\tau_*}{b} - \frac19 + \frac3{4 \; g} \right]
\eea
As a consequence, the scale factor near the singularity takes the form
\be
a(\tau) \buildrel{\tau \to \tau_* }\over= C \; (\tau-\tau_*)^\frac13 \left[1 + 
(\tau-\tau_*)^2 \; 
P_4^{a}\left(\log\frac{\tau-\tau_*}{b}\right) \right] \; .
\ee
where the coefficients of the fourth order polynomial $ P_4^{a} $ can be
obtained from eqs.(\ref{sing1}) and (\ref{expli}).

\subsection{Quantum loop effects and the validity of the classical inflaton picture}

When $ \tau \to \tau_* $ quantum loop corrections are expected to become very large 
spoiling the classical description. More precisely, quantum loop 
corrections are of the order $ (H/M_{Pl})^2 $ \cite{biblia}. From eqs.(\ref{tau}) and 
(\ref{sing1}) the quantum loop corrections are of the order
$$
\left(\frac{H}{M_{Pl}}\right)^2 \buildrel{(\tau-\tau_*) \ll 1}\over= 
\left[\frac{m}{3 \; (\tau-\tau_*) \; M_{Pl}}\right]^2 = 
\left(\frac{1.66 \; 10^{-6}}{\tau-\tau_*}\right)^2= \frac19 \; 
\left(\frac{\tau_{Planck}}{\tau-\tau_*}\right)^2
$$
where we used $ m = 1.21 \; 10^{13} $ GeV \cite{biblia}.

The characteristic time is here the Planck time
$$
\tau_{Planck} = m \; t_{Planck} = \frac{m}{M_{Pl}} = 2.703 \; 10^{-43} \; {\rm sec} \times m 
= 4.97 \; 10^{-6} \; .
$$
Namely, the  quantum loop corrections are less than 1\% for times
\be\label{corrQ}
(\tau-\tau_*) > \frac{10}3 \; \tau_{Planck}  = 1.66 \; 10^{-5} \; .
\ee
Therefore, for times $ (\tau-\tau_*) > 10^{-5} $ the classical treatment of the inflaton 
and the space-time presented in sec. \ref{uno} and \ref{persi} can be trusted and
we see that the classical description has a wide domain of validity.

The use of a classical and homogeneous inflaton field is justified
in the out of equilibrium field theory context as the quantum formation
of a condensate during inflation. This condensate turns to obey the classical evolution
equations of an homogeneous inflaton \cite{eri}. 

We see from eq.(\ref{sing1}) that the inflaton field becomes negative for 
$ \tau \to \tau_* $. But since a condensate field should be always positive, the
classical and homogeneous inflaton picture requires
$$ 
\tau-\tau_* > b 
$$ 
For the best fit coupling $ y = 1.26 $ and 63 efolds
of inflation we have $ b = 4.745272\ldots \; 10^{-5} = 9.55 \; \tau_{Planck} $
which is consistent with eq.(\ref{corrQ}).
By comparing this value of $ b $ with eq.(\ref{corrQ}) we see that the quantum loop 
corrections are negligible in the stage where the condensate is already formed.

We can obtain a lower bound on $ b $ since $ b $ increases with the 
number of inflation efolds $ N_{sr} $ at fixed inflaton potential
and since $ N_{sr} $ cannot be smaller than 
the lower bound provided by flatness and entropy \cite{biblia}.

\medskip

Although all inflationary solutions obtained evolving backwards in time from
the slow-roll stage do reach a zero of the scale factor, 
such mathematical singularity is attained extrapolating 
the classical treatment where it is {\bf no more valid}.
In fact, one never reaches the singularity in the validity region of the
classical treatment. 
In summary, the classical singularity at $ \tau=\tau_* $ 
{\bf is not a real physical phenomenon} here.

\medskip

The classical description with the homogeneous inflaton is very good for 
$ \tau-\tau_* > 10 \;\tau_{Planck} $ well before the beginning of inflation.

\subsection{The fast-roll regime: analytic approach}

As we see from fig. \ref{evolu} the inflaton field $ \phi(\tau) $ is much 
smaller than $ d\phi/d \tau $ during fast-roll. We can therefore 
approximate the coupled inflaton evolution equation and Friedmann equation 
eqs.(\ref{evol}) as
\bea \label{evfr} 
&& \frac{d^2 \phi}{d \tau^2} + 3 \; h \; \frac{d\phi}{d \tau} 
= 0 \quad , \cr \cr
&&  h^2(\tau) = 
\frac13\left[\frac12 \left(\frac{d\phi}{d \tau}\right)^2 + \frac1{4 \; g}
\right] \quad .
\eea 
Or, in a compact form,
\be
\frac{d^2 \phi}{d \tau^2} + \sqrt{\frac32} \; \frac{d\phi}{d \tau} \;
\sqrt{\left(\frac{d\phi}{d \tau}\right)^2 +  \frac1{2 \; g}}=0 \; ,
\ee
which has the exact solution
\be \label{solfr}
 \frac{d\phi}{d \tau} = \sqrt{\frac23} \; \frac1{\tau_1 \; 
\sinh\left(\displaystyle\frac{\tau-\tau_*}{\tau_1}\right)}
\quad , \quad   \phi(\tau) = \sqrt{\frac23} \; 
\log\left[\frac{2 \; \tau_1}{b} \;
\tanh\left(\frac{\tau-\tau_*}{2 \; \tau_1}\right)\right] \quad ,
\ee 
where $ \tau_1 $ turns out to be the characteristic time scale
\be\label{tau1}
\tau_1 =  2 \; \sqrt{\frac{g}3} = \sqrt{\frac{y}{6 \;N}} \; .
\ee
We find for the best fit to CMB and LSS data, $ y = 1.26 $ and $ N = 60 $,
\be\label{vtau1}
\tau_1 = 0.0592 = 11910 \; \tau_{Planck} \; ,
\ee
well after the Planck scale $ \tau_{Planck} = 4.97 \; 10^{-6} $.

The integration constant in eq.(\ref{solfr}) matches with the small 
$ \tau -\tau_* $ behaviour eq.(\ref{sing1}). The Hubble parameter and 
the scale factor are here 
\be \label{hafr}
 h(\tau) = \frac1{3 \; \tau_1} \; \coth u
\quad , \quad
 a(\tau) =  C \; \left[ \tau_1 \; \sinh u \right]^{\! \frac13} 
\quad , \quad u \equiv \frac{\tau-\tau_*}{\tau_1} \quad , 
\ee
where the integration constant was chosen to fulfil eq.(\ref{acero}).
The scale factor eq.(\ref{hafr}) interpolates between the non-inflationary
power law behaviour eq.(\ref{acero}) for $ \tau -\tau_* \to 0 $ and the eternal 
inflationary de Sitter behaviour for $ \tau -\tau_* \gg \tau_1 $. 
Since we have set $ v(\phi) $ equal to 
constant, slow-roll De Sitter inflation {\bf never stops} in this approximation.
Namely, neither matter-dominated nor radiation-dominated eras are reached
in this approximation.

\medskip

We can eliminate the variable $ u $ between $ \phi $ and $ d\phi/d \tau $ in eq.(\ref{solfr})
with the result
\be\label{fifip}
\frac{d\phi}{d \tau} = \sqrt{\frac23} \; \left[ \frac{e^{-\sqrt{\frac32} \; \phi(\tau)}}{b}
- \frac{b}{4 \; \tau_1^2} \; e^{ \sqrt{\frac32} \; \phi(\tau)} \right] \; .
\ee
This equation generalizes eq.(\ref{fifis}) which corresponds to 
the first term here and describes the behaviour for $ \tau -\tau_* $.
Notice that 
$$ 
-\infty < \phi(\tau) < \sqrt{\frac23} \; 
\log\left[\frac{2 \; \tau_1}{b}\right] \quad , \quad 0 < \frac{d\phi}{d \tau} < +\infty
$$
and that $ b /[2 \;\tau_1] = 4.0105 \; 10^{-4} $.

\medskip

The evolution described by eqs.(\ref{solfr})-(\ref{hafr}) starts from the 
mathematical singularity at $ \tau = \tau_* $ with monotonically decreasing 
$ d\phi/d \tau $ and $ h(\tau) $ and a monotonically increasing  
$ \phi(\tau) $ from its initial value $ \phi(\tau_*) = -\infty $.

\medskip

Slow-roll is reached asymptotically for large $ \tau $ since
$ d\phi/d \tau $ vanishes for $  \tau -\tau_* \to \infty $. 

\medskip

We find for the parameter $ \epsilon_v $ [eq. (\ref{epsv})] and for the equation of 
state,
\be\label{epsfr}
\epsilon_v(\tau) = \frac3{1 + \sinh^2 u}
 \quad , \quad  \frac{p(\tau)}{\rho(\tau)} = 
\frac2{\cosh^2 u} - 1 \quad .
\ee
We see that $ \epsilon_v(\tau) $ monotonically decreases with $ \tau $
and vanishes for $ \tau-\tau_* \to \infty $.
The equation of state $ p/\rho $ smoothly interpolates between $ + 1 $ at $ \tau = \tau_* $
(extreme non-inflationary fast-roll)
and $ - 1 $ (slow-roll inflation) for $ \tau - \tau_* \to \infty $, passing by
$ p/\rho = -1/3 $ (the beginning of fast-roll inflation) at $ \tau - \tau_* = 0.0573 $.

\medskip

The potential $ V_\mathcal{R}(\tau) $ eq.(\ref{fluces}) felt by the fluctuations 
takes here the form
\be\label{vafr}
V_\mathcal{R}(\tau) = \frac1{6 \; g} \left[1 - \frac1{2 \; \sinh^2 u}
- \frac9{\cosh^2 u} \right] \quad , \quad u = \frac{\tau-\tau_*}{\tau_1} \; .
\ee

\medskip

The limiting values of $ h(\tau), \; \phi(\tau) $ and 
$ V_\mathcal{R}(\tau) $ for $ \tau \to \infty $ give a reasonable 
approximation to the numerical results. We have
\be\label{infi}
h(\infty) = \frac1{3 \; \tau_1} =
 \sqrt{\frac{2 \; N}{3 \; y}} \quad , \quad
\phi(\infty) = \sqrt{\frac23} \; 
\log\left[\frac{2 \; \tau_1}{b} \right] \quad , \quad
 \frac{d\phi}{d\tau}(\infty) = 0 \quad , \quad
V_\mathcal{R}(\infty) = \frac{4\; N}{3 \; y} \; .
\ee
The characteristic time scale $ \tau_1 $ is generically a small number since
according to  eq.(\ref{tau1}) $ \tau_1 \sim 1/\sqrt{N} $. The value of  $ \tau_1 $
for the  best fit value for $ y $ is given in eq.(\ref{vtau1}).

\medskip

The end of fast-roll $ \tau_{trans} $ can be estimated in this approximation
by  using eq.(\ref{epsfr}) for  $ \epsilon_v(\tau) $
setting  $ \epsilon_v(\tau_{trans}) = 1/N $. This gives,
$$
\epsilon_v(\tau) \simeq 12 \; e^{-\frac{2 \; \tau_{trans}}{\tau_1}} 
= \frac1{N} \quad , \quad \tau_{trans} \simeq \frac12 \; \tau_1 \; 
\ln(12 \; N) = 0.195 \; .
$$
This approximated value for $ \tau_{trans} $ should be compared with the 
exact numerical result $ \tau_{trans} = 0.2487963\ldots $. $ h(\tau_{trans}) $ and 
$ V_\mathcal{R}(\tau_{trans}) $ differ in less than $ 1\% $
from their values at $ \tau = \infty $ given by eq.(\ref{infi}).

\medskip

In figs. \ref{apefi} we plot
$ \ln a(\tau), \; \ln h(\tau), \; \phi(\tau), \; \ln |{\dot \phi}(\tau)|,
\epsilon_v(\tau) $ and $ \; p(\tau)/\rho(\tau) $ computed
{\bf numerically} and computed using the analytic expressions
eqs.(\ref{solfr})-(\ref{epsfr}). We compare in figs. \ref{potas} 
the exact potential $ V_\mathcal{R}(\tau) $ with the analytic approximation eq.(\ref{vafr}).

We see that the simple analytic formulas
eqs.(\ref{solfr})-(\ref{vafr}) provide a very good approximation
during the fast-roll regime $ \tau \leq t_{trans} = 0.2487963\ldots $.
In particular,  eq.(\ref{solfr}) provides an excellent approximation to 
$ \phi(\tau) $ as shown in fig. \ref{apefi}.
In particular, the analytic formulas eqs.(\ref{solfr})-(\ref{vafr}) 
become {\bf exact} near the singularity at $ \tau = \tau_* $.

\subsection{The fast-roll regime: numerical solution}

To construct a singular solution we can integrate eqs.~\eqref{evol} 
backwards in time starting from initial conditions of strong non-inflationary fast--roll
type, namely
$$
K \equiv \frac{\dot\phi^2}{2 \, v(\phi)} \gg 1 \; ,
$$
producing a given total number $ N_{sr} $ of slow--roll inflationary
efolds. For instance, we start from some $ \phi $ and $ \dot\phi $ such that 
$ K = 10^4 $. The time extent backwards from this moment
has to be limited so that, integrating back and forth, the required relative accuracy of
$ 10^{-12} $ is preserved. We furthermore impose that $ N_{sr}=63 $. 

\medskip

We adopt the convention that conformal time $ \eta $ vanishes from below 
when inflation ends and that $ a(\tau=0)=1 $ when there are still $ N=60 $ 
efolds till the end of inflation. This choice of the scale factor normalization 
seems the most natural.
Then, $ \eta $ has a finite non-zero limit $ \eta_\ast $ as 
$ \tau $ approaches the time $ \tau_\ast $ of the singularity, 
since $ a( \tau)\simeq C \; (\tau- \tau_\ast)^{1/3} $ as $ \tau\to \tau_\ast $ 
according to eq.(\ref{acero}). That is,
$$
\eta = \int_{\tau_{end}}^\tau \frac{d\tau'}{a(\tau')}=
\eta_\ast + \int_{\tau_\ast}^\tau \frac{d\tau'}{a(\tau')} \; .
$$
The numerics of a fast--roll solution of this type are in Table I
where a relative accuracy of $ 10^{-12} $ is preserved.

Using the asymptotic behaviour eq.(\ref{sing1}) as $ \tau\to \tau_\ast^+ $ 
we obtain from Table I: 
$$
\tau_\ast = -0.8499574\ldots \quad ,  \quad b = 4.745272\ldots
10^{-5} \quad {\rm and} \quad  \eta_\ast = -15.605614\ldots \; .
$$
Slow--roll begins at
$ \tau_{trans} = \tau_\ast+0.2487963\ldots = -0.6011611\ldots$.

\medskip

The initial value of the ratio
$$ 
\displaystyle \frac{d\varphi/dt}{\varphi} = m \; \frac{\dot\phi}{\phi} 
$$ 
has the dimension of mass. The natural mass scale in the problem
is here the energy scale of inflation $ M $. Therefore, assuming this ratio of the
order $ M $ yields
$$
\frac{\dot\phi}{\phi} < \frac{M}{m} \sim 10^3 \; .
$$
Hence, it s natural to  
start the fast--roll evolution with $ \dot\phi/\phi < 10^3 $.

\begin{table}[h]
   \begin{tabular}{|c|c|c|c|c|c|c|}
\hline
&  $ K=5.3458 \ldots \; 10^7$ & $ K=10^4 $ & inflation start: $ \ddot a=0^-$
& fast-roll $\to$ slow-roll & $ a=1 $  & inflation end: $ \ddot 
a=0^+$ \\
\hline \hline
$ \tau $ &  $-0.8499493\ldots$ & $-0.8493593\ldots$ & $-0.7746494\ldots$
& $-0.6011611\ldots$ &  0  &  $17.4048242\ldots$  \\
\hline
$ \phi $ & $ -1.4401237\ldots$  & $2.0690604\ldots$ & $5.9342489\ldots$
& $6.4783577\ldots$& $6.7484076\ldots$&  $18.5586530\ldots$ \\
\hline
$ \dot\phi $ & $100391.035\ldots$ & $1365.05241\ldots$ & $8.8601670\ldots$
& $0.9182661\ldots$& $0.3974015\ldots$& $0.94150557\ldots$ \\
\hline
$ \log a $ & $-7.0325621\ldots$ & $-5.5999353\ldots$ & $-3.9142151\ldots$
& $-2.9999999\ldots$&  $0$   & $60$    \\
\hline
$h$ & $40984.4689\ldots$& $557.30817\ldots$ & $6.2650841\ldots$
& $5.0295509\ldots$& $4.9653990\ldots$& $0.6657449\ldots$   \\
\hline
$ \eta $ & $-15.6050091\ldots$  & $-15.376218\ldots$& $-15.3549996\ldots$
& $-4.0169827\ldots$& $-0.2020609\ldots$ & $0$    \\
\hline
   \end{tabular}
\caption{Fast-roll solution with $ N_{sr} = 63 $ efolds of slow-roll
inflation. Recall that $ \tau = 4.97 \; 10^{-6} \; (t/t_{Planck}) $.}
\end{table}

\section{The slow-roll inflationary era}

\subsection{The extreme slow-roll solution}\label{esr}

There always exist a special solution of eqs.(\ref{evol}) that starts
at $ \tau = -\infty $ with vanishing inflaton, vanishing scale factor
but nonzero Hubble parameter. More precisely, eqs.(\ref{evol}) can
be approximated for small $ \phi $ and $ \dot \phi $ as
\bea \label{evola} 
&& \frac{d^2 \phi}{d \tau^2} + 3 \; h \; \frac{d\phi}{d \tau} -\phi = 0 
\quad , \cr \cr
&&  h^2(\tau) = \frac13 \; v(0) \quad .
\eea 
where we used eqs.(\ref{vpol}) and (\ref{evol}).

Eqs.(\ref{evola}) admit the asymptotic solution for  $ \tau \to -\infty $
\be\label{asiex}
\phi(\tau) \buildrel{\tau \to -\infty  }\over= C_0 \; 
e^{\alpha \; \tau} \to 0
\quad , \quad  h(\tau)\buildrel{\tau \to -\infty  }\over= 
\sqrt{\frac{v(0)}3}
\quad , \quad  a(\tau)\buildrel{\tau \to -\infty  }\over= 
e^{\sqrt{\frac{v(0)}3} \; \tau} \to 0 \; ,
\ee
where  $ C_0 $ is an integration constant, $ v(0) = 2 \, N/y $ for the double-well 
potential eq.(\ref{vnue}) and
$$
\alpha \equiv \frac12 \left[ \sqrt{3 \; v(0)+4}-
\sqrt{3 \; v(0)}\right] > 0 \; .
$$
Notice that $ \alpha $ can be expressed in terms of the fast-roll characteristic time-scale
$ \tau_1 $ [eq.(\ref{tau1})],
$$
\alpha = \frac1{2 \; \tau_1}\left[ \sqrt{1 + 4 \; \tau_1^2} - 1 \right] \simeq \tau_1
$$
since $ \tau_1 \simeq  0.0592 \ll 1 $ [see eq.(\ref{vtau1})].

\medskip

It must be noticed that the characteristic time scale of the inflaton evolution
in the extreme slow-roll solution for early times [see eq.(\ref{asiex})]
$$ 
\frac1{\alpha} \simeq \frac1{\tau_1} \gg 1 \; ,
$$
turns to be the {\bf inverse} of the 
characteristic time scale $ \tau_1 $ of the fast-roll solution and to be very large. 

On the contrary, the characteristic time scale of the scale factor evolution in the same regime is
very short
$$
\sqrt{\frac3{v(0)}} = 3 \; \tau_1 \ll 1 \; .
$$
The fast-roll stages both non-inflationary and inflationary are absent in this solution. 
The extreme slow-roll solution only possesses the slow-roll inflationary stage followed by the 
matter dominated era. 

\begin{table}[h]
  \begin{tabular}{|c|c|c|c|c|}
\hline  
    & inflation start & $ a=1 $  &  inflation end: $ \ddot a=0^+ $ \\
\hline \hline 
   $ \tau $      & $-344.9514017\ldots$&  0    & $17.40482446\ldots$  \\
\hline
   $ \phi $ & $ 10^{-8} $ &  $ 6.7484118\ldots$  & $ 18.5586530\ldots$  \\
\hline
$ \dot\phi $ & $\alpha\,10^{-8}=5.8937108453...10^{-10}$ & 
$0.3973384\ldots$ & $0.94150557\ldots$\\
\hline 
   $ \log a $ &$-1938.4867948\ldots$&  0   &   60     \\
\hline 
   $ h $ &  $ \sqrt{2 \, N/(3 \, y)} = 
5.6361006\ldots$   &  $4.9653973\ldots$ & $0.6657449\ldots$  \\
\hline
$ \eta $ &  $-\infty~$(f.a.p.p) & $-0.2020610\ldots$ & 0    \\
\hline
  \end{tabular}
  \caption{Relevant quantities of the extreme slow-roll inflaton solution
for the coupling $ y = 1.2592226\ldots$.
We adopt the convention that $ a(\tau=0)=1 $ when there are still $ N=60 $ 
efolds till the end of inflation. Recall that $ \tau = 4.97 \; 10^{-6} \; (t/t_{Planck}) $.} 
\end{table}

For the value of the coupling $ y = 1.2592226\ldots$, we get
for the extreme slow--roll solution
\be\label{eq:phend}
  \alpha = 0.058937108\ldots \;,\quad
  \phi_{\rm end} = 18.5586530\ldots \; ,\quad
  \dot\phi_{\rm end}  =  0.9415055\ldots
\ee
In table II we display the values of the relevant magnitudes for this
extreme slow-roll solution.

\medskip

Except for the extreme slow--roll solution, 
all solutions are of fast-roll type and come from singular values of $ \phi $ and $ h $ 
according to eq.(\ref{sing1}) as $ \tau\to \tau_\ast^+ $ 
for some finite  $ \tau_\ast $ characteristic of each particular
solution. The slow-roll stage 
(which starts when $ \epv=1/N $ from above, and ends when 
again $ \epv=1/N $ from below)
of all distinct solutions turns to be almost identical to that of the 
extreme slow--roll case as one could expect for an attractor.

\subsection{The inflaton during slow-roll inflation: analytical solution}\label{solusr}

In the slow-roll regime higher time derivatives can be neglected in the
evolution eqs.(\ref{evol}) with the result
\be\label{sr1}
3 \, h(\tau) \; {\dot \phi} + v'(\phi) = 0 \quad ,  \quad 
h^2(\tau) = \frac{v(\phi)}3 \; .
\ee
These first order equations can be solved in closed form as
\be \label{Nefo}
N[\phi] = -\int_{\phi}^{\phi_{end}}  \;
v(\phi') \; \frac{d \phi'}{dv} \; d\phi' \;  \; .
\ee
where $ N[\phi] $ is the number of e-folds since the field $ \phi $ exits the 
horizon till the end of inflation (where it takes the value $ \phi_{end} $). 

Eq.(\ref{Nefo}) indicates that $ N[\phi] $ scales as $ \phi^2 $ and 
hence the field $ \phi $ is of the order $ \sqrt{N} \sim \sqrt{60} $. 
Therefore, we proposed as universal form for the inflaton potential 
\cite{1sN,biblia}
\be \label{V} 
v(\varphi) = N \; M^4 \; w(\chi)  \; ,
\ee  
\noindent where $ \chi $ is the dimensionless, slowly varying field 
\be\label{chifla} 
\chi = \frac{\varphi}{\sqrt{N} \;  M_{Pl}} = \frac{\phi}{\sqrt{N}} \; .
\ee 
The equations of motion (\ref{evol}) in the field $ \chi $ become
\bea \label{evol2} 
&&  {\cal H}^2({\hat \tau}) = \frac13\left[\frac1{2\;N} 
\left(\frac{d\chi}{d {\hat \tau}}\right)^2 + w(\chi) \right] \quad  
{\rm with} \quad {\cal H} = \frac{h}{\sqrt{N}} \; , \cr \cr
&& \frac1{N} \;  \frac{d^2
\chi}{d {\hat \tau}^2} + 3 \;  {\cal H} \; \frac{d\chi}{d {\hat \tau}} + w'(\chi) = 0 
\quad .
\eea 
and $ \hat \tau $ stands for the rescaled dimensionless time
$$
{\hat \tau} \equiv \frac{\tau}{\sqrt{N}} = \frac{m \; t}{\sqrt{N}} \; .
$$
To leading order in the slow-roll approximation (neglecting $ 1/N $ 
corrections), eqs.(\ref{evol2}) are solvable in terms of quadratures
\be\label{tausr}
{\hat \tau} -{\hat \tau}_{trans} = - \int_{\chi({\hat \tau}_{trans})}^\chi d\chi' \; 
\frac{\sqrt{3 \; w(\chi')}}{ w'(\chi')}\; ,
\ee
where $ {\hat \tau}_{trans} $ stands for the beginning of slow-roll inflation 
and we used that 
\be\label{hsr}
{\cal H}({\hat \tau}) = \sqrt{\frac{w(\chi)}3} + 
{\cal O}\left(\frac1{N}\right) \; ,
\ee
For the broken symmetric potential eq.(\ref{bini}), from eqs.(\ref{enerpres}), 
(\ref{tausr}) and (\ref{hsr}), we find
\bea\label{chislr}
&& \chi({\hat \tau}) = \chi({\hat \tau}_{trans}) \; e^{\sqrt{\frac{y}6} \; 
({\hat \tau}-{\hat \tau}_{trans})} 
+{\cal O}\left(\frac1{N} \right)
= \sqrt{\frac8{y}} \;  e^{-\sqrt{\frac{y}6} \; ({\hat \tau}_{end}-{\hat \tau})} 
+{\cal O}\left(\frac1{N} \right)
\; , \\ \cr
&& {\cal H}({\hat \tau}) =\sqrt{\frac2{3 \; y}}
\left[ 1 -  e^{-\sqrt{\frac{2 \; y}3} \; ({\hat \tau}_{end}-{\hat \tau})}
\right] +{\cal O}\left(\frac1{N} \right) \; , \cr \cr
&& \frac{p}{\rho}({\hat \tau}) = -1 + \frac{y}{6 \; N} \; 
\frac1{\sinh^2\left[\sqrt{\frac{y}6}({\hat \tau}_{end}-{\hat \tau})\right]} 
+{\cal O}\left(\frac1{N^2} \right)\; , \label{ecesta} \\ \cr
&& {\rm for} \quad {\hat \tau}_{trans} \leq {\hat \tau} \leq {\hat \tau}_{end} = \sqrt{\frac3{2 \; y}} \; 
\ln \left[\frac8{\chi^2({\hat \tau}_{trans}) \; y}\right] + 
{\cal O}\left(\frac1{\sqrt{N}} \right) \label{taufin} \; .
\eea
Inflation ends when the equation of state becomes 
$ p/\rho = -1/3 $ [see eq.(\ref{aseg})]. 
According to eq.(\ref{ecesta}), this happens when
$ {\hat \tau}_{end} -{\hat \tau} \sim {\cal O}\left(1/\sqrt{N} \right) $. Therefore,
expressions eqs.(\ref{chislr})-(\ref{ecesta}) are valid as long as
$$
{\hat \tau}_{trans} \leq {\hat \tau} \leq {\hat \tau}_{end} -  {\cal O}\left(\frac1{\sqrt{N}} \right) 
\quad {\rm where} \quad   {\cal O}\left(\frac1{\sqrt{N}}\right)  > 0 \; .
$$
That is, eqs.(\ref{chislr}) hold while the inflaton is not very near the
minimum of the potential $ \chi_{end} = \sqrt{8/y} $.

\medskip

By integrating the Hubble parameter $ {\cal H}({\hat \tau}) $
we obtain for the scale factor $ a({\hat \tau}) $ 
\bea \label{asr}
&& \log \frac{a({\hat \tau})}{a({\hat \tau}_{trans})} = \sqrt{\frac2{3 \; y}} \; N \; 
({\hat \tau} -{\hat \tau}_{trans})-
\frac{N}8 \; \chi^2({\hat \tau}_{trans}) \left[e^{\sqrt{\frac{2 \; y}3} \; 
({\hat \tau}-{\hat \tau}_{trans})} - 1 \right] =\\ \cr
&&= \sqrt{\frac{2 \; N}{3 \; y}} \; m \; (t -t_{trans})-\frac18 \; 
\left[\frac{\varphi(t_{trans})}{M_{Pl}}\right]^2 \; 
\left[e^{\sqrt{\frac{2 \; y}{3 \; N}} \; m \; (t-t_{trans})} - 1 \right] \; ,
\nonumber
\eea
where we used eqs.(\ref{tau} ) and (\ref{chislr}). It must be noticed
that $ a({\hat \tau}) $ is not exactly a de Sitter scale factor, even in the large 
$ N $ limit at fixed $ {\hat \tau} $.

At the end of inflation the number of efolds is $ \ln a \simeq 
64 $, the inflaton is near its minimum 
$$ 
\chi = \sqrt{\frac8{y}} \simeq 2.52 \quad , 
$$ 
$ \dot \chi $ starts to oscillate around zero and $ \cal{H}({\hat \tau}) $ 
begins a rapid decrease (see figs. \ref{evolu}). At this time the inflaton field 
is no longer slowly coasting in the $ w''(\chi) < 0 $ region but rapidly 
approaching its equilibrium minimum. When inflation ends, the inflaton is 
at its minimum value up to corrections of order $ 1/\sqrt{N} $. Therefore, we 
see from the Friedmann eq.(\ref{evol2}) and eqs.(\ref{chislr}) that
\be\label{reduH}
\frac1{N} \left(\frac{d\chi}{d {\hat \tau}}\right)^{\! 2}({\hat \tau}_{end}) = 
{\cal O}\left(\frac1{N}\right) \quad , \quad 
w(\chi({\hat \tau}_{end})) = {\cal O}\left(\frac1{N}\right) \quad 
{\rm and ~~ therefore,} \quad {\cal H}({\hat \tau}_{end}) = 
{\cal O}\left(\frac1{\sqrt{N}} \right) \; ,
\ee
while $ {\cal H}({\hat \tau}_{trans}) = {\cal O}(1) $. Namely, the Hubble parameter 
decreases by a factor of the order $ \sqrt{N} \sim 8 $ during slow-roll 
inflation. We see in fig.  \ref{evolu} that the exact $ \cal{H}({\hat \tau}) $ decreases 
by a factor six during slow-roll inflation,
confirming the slow-roll analytic estimate.

We can compute the total number of inflation efolds $ N_{tot} $ to leading
order in slow-roll inserting the analytic formula for $ {\hat \tau}_{end} $ 
eq.(\ref{taufin}) in eq.(\ref{asr}) with the result,
\be\label{ntotsr}
N_{tot} = \frac{N}{y} \left\{\ln\left[ \frac8{\chi^2({\hat \tau}_{trans}) \; y}\right] - 1 + 
\frac18 \; y \; \chi^2({\hat \tau}_{trans}) \right\} + 
{\cal O}\left(\frac1{\sqrt{N}} \right) \; .
\ee
We have verified the slow-roll analytical results 
eqs.(\ref{chislr})-(\ref{ntotsr}) comparing them with the numerical 
solution of eqs.(\ref{evol}). Both results are concordant up to the error 
estimation in each case: $ {\cal O}\left(1/N \right) $ or 
$ {\cal O}\left(1/\sqrt{N} \right) $.

\medskip

The field $ \phi $ as a function of the dimensionless time $ \tau $
eq.(\ref{chislr}) takes the form
$$
\phi(\tau) = \phi({\tau}_{trans}) \; e^{\sqrt{\frac{y}{6 \; N}} \; 
(\tau-\tau_{trans})} 
$$
and then
$$
{\dot \phi(\tau)} =\sqrt{\frac{y}{6 \; N}} \; \phi(\tau) \; .
$$
For $ y \simeq 1.26 $ and $ N = 60 $ we get $ \sqrt{y/[6 \; N]} = 0.0577 $
in agreement with the slope of the red quasi-horizontal slow-roll line in the phase
space flow fig. \ref{phs}.

\section{Complete Fluctuations evolution and fast-roll 
effects on the power spectrum.}

\subsection{Scalar and tensor fluctuations near the initial singularity.}

In order to study the curvature and tensor fluctuations in this regime, 
it is important to evaluate the parameter $ \epsilon_v $ and the potential 
felt by the fluctuations $ V_\mathcal{R} $.

Inserting eqs.(\ref{sing2}) and (\ref{defP}) into eqs.(\ref{epsv}) and 
(\ref{wtau}) yields near the initial singularity
\bea\label{wep0}
&& V_\mathcal{R}(\tau)\buildrel{\tau \to \tau_* }\over=
- \frac1{9 \; (\tau-\tau_*)^2} \left[1 +  (\tau-\tau_*)^2 \; 
P_4^{V}\left(\log\frac{\tau-\tau_*}{b}\right) \right] \quad , \quad 
\epsilon_v \buildrel{\tau \to \tau_* }\over= 3 \left[1 +  (\tau-\tau_*)^2 \; 
P_4^{\epsilon}\left(\log\frac{\tau-\tau_*}{b}\right) \right] \\ \cr
&& 
W_\mathcal{R}(\eta)\buildrel{\eta \to 0 }\over=-\frac1{4 \;  \eta^2}
\left[1 +  \eta^3 \; P_4^{W}\left(\log \eta\right)\right] \; .
\eea
where  
$$ 
\eta  \buildrel{\tau \to \tau_* }\over= \frac32 \; (\tau-\tau_*)^\frac23 
$$ 
is the conformal time for 
$ \tau \to \tau_* $ and $ P_4^{V}(x), \;  P_4^{\epsilon}(x) $ and $ P_4^{W}(x) $
are polynomials of degree four in $ x $ .

We see that the fluctuations feel a {\bf singular attractive} potential
near the $ \eta = 0 $ singularity. Actually, the behaviour of 
$ W_\mathcal{R}(\eta) $ for $ \eta \to 0 $ is {\bf exactly} the 
{\bf critical} strength ($ -1/4 $) for which the fall to the centre becomes possible in a 
central and attractive singular potential \cite{ll}. 

We find from eqs.(\ref{flutau}) and (\ref{wep0}) for the fluctuations near 
the singularity 
\be\label{sr0}
S_\mathcal{R}(k;\eta)  \buildrel{\eta , \eta_0 \to 0 }\over= \sqrt{\frac{\eta}{\eta_0}}
\left[ {\cal A}_\mathcal{R}(k) +  {\cal B}_\mathcal{R}(k)
 \; \log \frac{\eta}{\eta_0} \right] \; ,
\ee
where $ \eta_0 $ is the time when the initial conditions will be imposed, 
$ {\cal A}_\mathcal{R} $ and $ {\cal B}_\mathcal{R} $ are complex constants constrained by the 
Wronskian condition (that ensures the canonical commutation relations)
\cite{biblia}
\be\label{wronskian}
W[S_\mathcal{R},S^*_\mathcal{R}]= S_\mathcal{R} \; 
\frac{dS^{*}_\mathcal{R}}{d \eta} - 
\frac{dS_\mathcal{R}}{d \eta} \; S^*_\mathcal{R} = i \; .
\ee
Namely,
\be\label{conwr}
2 \; {\rm Im}[ {\cal A}_\mathcal{R} \;  {\cal B}_\mathcal{R}^* ] = \eta_0 \; .
\ee
Precisely, the logarithmic behaviour for $ \eta \to 0 $ of the wave 
function eq.(\ref{sr0}) describes the fall to $ \tau - \tau_* = 0 $ for the 
critical strength of the potential $ W_\mathcal{R}(\eta) $. 
For larger attractive strengths the wave function eq.(\ref{sr0})
shows up an oscillatory behaviour \cite{ll}. Notice, however the physical
nature of the process:
here we have a time evolution near a classical singularity
at a given time while in the potential case one has particles falling
(or emerging) from a point in space where the potential is singular.

\medskip

In general, the mode functions for large $ k $ must behave as 
free modes (plane waves) since the potential $ W_\mathcal{R}(\eta) $ 
in eq.(\ref{fluces}) becomes negligible in this limit except at the singularity
$ \tau=\tau_* $. One can then impose
Bunch-Davies conditions for large $ k $ which corresponds to
assume an initial quantum vacuum Fock state, empty of curvature excitations
\cite{biblia}
\be\label{BDk}
S_\mathcal{R}(k;\tau) \buildrel{k \to \infty }\over= 
\frac{e^{-i \; k \; \eta}}{\sqrt{2 \; k}}
\ee
and therefore
$$
\frac{dS_\mathcal{R}}{d \eta}(k;\eta_0) \buildrel{k \to \infty }\over= 
-i \; k \; S_\mathcal{R}(k;\eta_0)\; .
$$
Eq.(\ref{BDk}) fulfils the Wronskian normalization eq.(\ref{wronskian}).

In asymptotically flat (or conformally flat) regions of the space-time
the potential felt by the fluctuations vanish and the fluctuations
exhibit a plane wave behaviour for {\bf all} $ k $ (not necesarily large).
This is not the case near strong gravity fields or curvature singularities
as in the present cosmological space-time where $ W_\mathcal{R}(\eta) $ can 
never be neglected at fixed $ k $. However, we can choose Bunch-Davies initial 
conditions (BDic) at $ \eta = \eta_0 $ by imposing
\be\label{BD}
\frac{dS_\mathcal{R}}{d \eta}(k;\eta_0) = -i \; k \; 
S_\mathcal{R}(k;\eta_0) \quad {\rm for ~ all} \; k \; .
\ee
That is, we consider the initial value problem for the mode functions 
giving the values of $ S_\mathcal{R}(k;\eta) $ and 
$ dS_\mathcal{R}/d \eta $ at $ \eta = \eta_0 $.

Notice that eq.(\ref{BD}) combined with the Wronskian condition
eq.(\ref{wronskian}) implies that
$$
|S_\mathcal{R}(k;\eta_0)| = \frac1{\sqrt{2 \; k}} \quad , \quad
\left| \frac{dS_\mathcal{R}}{d \eta}(k;\eta_0) \right| = 
\sqrt{\frac{k}2} \; .
$$
which is equivalent to eq.(\ref{BDk}) for large $ k $.

\medskip

Since the mode functions $ S_\mathcal{R}(k;\eta) $ are defined up to
an arbitrary constant phase we can write eq.(\ref{sr0}) valid
near the metric singularity as 
\be\label{S}
S_\mathcal{R}(k;\eta)  \buildrel{\eta , \eta_0 \to 0 }\over= 
\sqrt{\frac{\eta}{2 \; k \; \eta_0}}\left[1 - \left(\frac12 + i \; k \; 
\eta_0\right) \; \log \frac{\eta}{\eta_0} \right] \; .
\ee
In eq.(\ref{sr0}) this corresponds to the coefficients,
$$
 {\cal A}_\mathcal{R}  = \frac1{\sqrt{2 \; k}} \quad ,  \quad  
{\cal B}_\mathcal{R} = \frac1{\sqrt{2 \; k}}
\left(\frac12 + i \; k \; \eta_0\right) \quad .
$$

We have in cosmic time,
\be\label{Stau}
S_\mathcal{R}(k;\tau) \buildrel{\tau , \tau_0 \to \tau_* }\over= 
\frac1{\sqrt{2 \; k}} \;
\left(\frac{\tau-\tau_*}{\tau_0}\right)^\frac13 
\left[1 - \left(\frac13 + i \; k \; 
\tau_0^\frac23 \right) \; \log \frac{\tau-\tau_*}{\tau_0} \right] \; ,
\ee
where $ \tau_0 - \tau_* = \left(2 \; \eta_0/3\right)^\frac32 $
for  $ \tau \to \tau_* $.

Namely, imposing the BD initial condition (BDic) eq.(\ref{BD}) at small
$ \eta_0 $ where the small $ \eta $ behavior eq.(\ref{sr0}) applies,
yields specific values for the coefficients of the linearly
independent solutions $ \sqrt{\eta} $ and  $ \sqrt{\eta} \, \log\eta $
that we can read from eqs.(\ref{S})-(\ref{Stau}).

\medskip

For general $ \tau_0 $ (i. e.,  $ \tau_0 $ not near $ \tau_* $), the mode functions for 
$ \tau \to \tau_* $ take the form
\be
S_\mathcal{R}(k;\tau) \buildrel{\tau \to \tau_* }\over= 
\frac1{\sqrt{2 \; k}} \;
\left(\frac{\tau-\tau_*}{\tau_0}\right)^\frac13 
\left[X(k,\tau_0) - \left(Y(k,\tau_0) + \frac{i \; k \; 
\tau_0^\frac23}{X(k,\tau_0)}\right) \log \frac{\tau-\tau_*}{\tau_0}
\right] \; ,
\ee
where we imposed eq.(\ref{conwr}) and we have from eq.(\ref{Stau}),
$$ 
X(k,\tau_*) = 1  \quad {\rm and}  \quad Y(k,\tau_*) = \frac13 \quad .
$$

Notice that $ X(k,\tau_0) > 0 $ for $ \tau_0 \to \tau_* $ as we see from
eq.(\ref{Stau}). Our numerical calculations show that $ X(k,\tau_0) > 0 $
{\bf for all} $ \tau_0 $ and $ k $.

\subsection{The primordial power spectrum, 
Scalar curvature fluctuations and the CMB+LSS data.}\label{priflu}

\begin{table}
\begin{tabular}{|c|c|c|}
\hline
Inflation (fast-roll) starts & $ {\ddot a}(\tau_s) = 0 $ & $ \tau_s = \tau_* + 0.0753090 $ \\ 
\hline $ V_\mathcal{R}(\tau) $ becomes positive & $ 
V_\mathcal{R}(\tau_+) = 0 $ & $ \tau_+ = \tau_* + 0.114 $ \\
\hline  End of fast-roll  &  $ N \; \epsilon_v(\tau_{trans}) = 1 $ &  
$ \tau_{trans} = \tau_* +0.2487963\ldots $ \\
\hline Maximum of $ V_\mathcal{R}(\tau) $ & $ V_\mathcal{R}'(\tau_M) = 0 $
& $ \tau_M = \tau_* +0.3503 \; , \; V_\mathcal{R}(\tau_M) = 51.196 $ \\ 
\hline  End of Inflation & $ p(\tau_{end}) = 0 $ & 
$ \tau_{end} = \tau_* + 18.2547816 $ \\ \hline
\end{tabular} 
\caption{Selected time values for $ y = 1.26 $ and $ N_{sr} = 63 $ efolds 
of slow-roll inflation. Notice that slow-roll starts exactly when fast-roll ends.
 Recall that $ \tau = 4.97 \; 10^{-6} \; (t/t_{Planck}) $.}
\end{table}

The power spectrum of curvature perturbations $ \mathcal{R} $ is given by the
expectation value $ <\mathcal{R}^2> $ in the state with general initial conditions
\cite{biblia}
\be\label{pot1}
< \mathcal{R}^2(\vx,\eta)> = \displaystyle \left(\frac{m}{M_{PL}}\right)^2
\int_0^{\infty} \frac{\displaystyle |S_\mathcal{R}(k;\eta)|^2}{\displaystyle z^2(\eta)} \; 
\frac{k^2 \, dk}{2 \, \pi^2} \; .
\ee
where $ z(\eta) $ is given by eq.(\ref{defwz}).
Notice in eq.(\ref{pot1}) the factor $ (m/M_{PL})^2 $ in the physical power 
spectrum expressed in terms of the dimensionless quantities used here.

The power spectrum at time $ \eta $ is customary defined as
the power per unit logarithmic interval in $ k $
$$
<\mathcal{R}^2(\vx,\eta)> =\int_0^{\infty} \frac{dk}{k} \; P_\mathcal{R}(k,\eta) \; .
$$
Therefore, the scalar power for general initial conditions 
is given by the fluctuations behavior by the end of 
inflation \cite{biblia} ,
\be\label{curvapot}
P_\mathcal{R}(k) = \left(\frac{m}{M_{PL}}\right)^2
\; \frac{k^3}{2 \; \pi^2} \; {\displaystyle \lim_{\eta \to  0^-}}
 \left|\frac{S_\mathcal{R}(k;\eta)}{z(\eta)} \right|^2 \; .
\ee 

The mode functions $ S_\mathcal{R}(k;\eta) $ obey the fluctuations equation (\ref{fluces})
where the potential  $ W_\mathcal{R}(\eta) $ [eq.(\ref{wtau})]
during slow-roll and to leading order in $ 1/N $ takes the simple form \cite{biblia} ,
\be \label{eqnz}
W_\mathcal{R}(\eta)= \frac2{\eta^2} \left[1+ \frac32\; (3
\; \epsilon_v-\eta_v) \right] = \frac{\nu^2_\mathcal{R}-
\frac14}{\eta^2} \quad , \quad \nu_\mathcal{R} = \frac32+ 3 \,
\epsilon_v -\eta_v +  {\cal O}\left(\frac1{N^2}\right) \; . 
\ee 
In the slow-roll regime we can consider $ \epsilon_v $ and $ \eta_v $ [see eq.(\ref{epsv})]
constants in time in eq.(\ref{eqnz}). During slow-roll,
the general solution of eq.(\ref{fluces}) is then given by 
\be \label{bogo}
S_\mathcal{R}(k;\eta) = A_\mathcal{R}(k) \;
g_{\nu_\mathcal{R}}(k;\eta) + B_\mathcal{R}(k) \; g^*_{\nu_\mathcal{R}}(k;\eta) \; , 
\ee 
with
\be\label{gnu}
g_{\nu }(k;\eta)  =  \frac12 \; i^{\nu +\frac12} \;
\sqrt{-\pi \eta} \; H^{(1)}_{\nu }(-k \; \eta) \; , 
\ee
$ A_\mathcal{R}(k), \;  B_\mathcal{R}(k) $ are constants determined by the 
initial conditions and $ H^{(1)}_{\nu}(z) $ is a Hankel function.

The Wronskian of the solutions $ S_\mathcal{R}, \; S_\mathcal{R}^* $ is given by 
eq.(\ref{wronskian}) and
$$
W[g_{\nu},g_{\nu}^*]=i 
$$
This generically determines that
\be\label{abc}
 |A_\mathcal{R}(k)|^2 - | B_\mathcal{R}(k)|^2 = 1 \; .
\ee

For wavevectors deep inside the Hubble radius $ | k \, \eta | \gg 1
$ the mode functions $ g_{\nu}(k;\eta) $ have the asymptotic behavior
\be
g_{\nu}(k;\eta) \buildrel{\eta \to
-\infty}\over= \frac1{\sqrt{2 \, k}} \; e^{-i \; k \; \eta} \quad ,  \quad
g^*_{\nu}(k;\eta) \buildrel{\eta \to -\infty}\over=
\frac1{\sqrt{2k}} \; e^{ i \; k \; \eta} \; , \label{fnuasy}
\ee
while for $ \eta \to 0^- $, they behave as:
\be\label{geta0}
g_\nu(k;\eta)\buildrel{\eta \to 0^-}\over=
\frac{\Gamma(\nu)}{\sqrt{2 \, \pi \; k}} \; \left(\frac2{i \; k \;
\eta} \right)^{\nu - \frac12} \; .
\ee
In particular, in the scale invariant case $ \nu=\frac32 $ which is
the leading order in the slow-roll expansion, the mode functions 
eqs.(\ref{gnu}) simplify to
\be\label{g32}
g_{\frac32}(k;\eta) =
\frac{e^{-i \; k \;\eta}}{\sqrt{2k}}\left[1-\frac{i}{k \;\eta}\right]\; .
\ee
As we see from eq.(\ref{defwz}), $ z(\eta) $ obeys eq.(\ref{fluces}) for $ k = 0 $ 
and therefore $ z(\eta) $ in the slow-roll regime behaves  as 
\be\label{zeta2} 
z(\eta) = \frac{z_0}{ (-k_0 \; \eta)^{\nu_\mathcal{R}-\frac12}} \; , 
\ee 
where $ z_0 $ is the value of $ z(\eta) $ when the pivot scale $ k_0 $ exits the horizon,
that is at $ \eta = -1/k_0 $. Combining this
result with the small $ \eta $ limit eq.(\ref{geta0}) we find from
eqs.(\ref{curvapot}) and (\ref{zeta2}), 
\be \label{powR}
P_{\mathcal{R}}(k) = P^{BD}_{\mathcal{R}}(k)\left[1+ D(k) \right] \; ,
\ee 
where we introduced the transfer function for the initial
conditions of curvature perturbations: 
\be \label{DofkR}
D(k) = 2 \; | {B}_\mathcal{R}(k)|^2 -2 \;
\mathrm{Re}\left[A_\mathcal{R}(k) \; B^*_\mathcal{R}(k)\,i^{2\nu_\mathcal{R}-3}\right] \; .
\ee
$ D(k) $ is obtained imposing BDic at $ \tau = \tau_0 $ according
to eq.(\ref{BD}). 

\medskip

Notice as shown in sec. \ref{dcero} that the transfer function $ D(k) $ enjoys the properties
\be\label{propg}
1 + D(k) \buildrel{k \to 0 }\over= {\cal O}(k^{n_s+1}) \quad , \quad
D(k) \buildrel{k \to \infty }\over= {\cal O}\left(\frac1{k^2}\right) \; .
\ee
$ D(k) $ accounts for the effect in the power spectrum both of the initial conditions 
and of the fluctuations evolution during fast-roll (before slow-roll).
$ D(k) $ depends on the time $ \tau_0 $ at which BDic are imposed.

\medskip

If one chooses the extreme slow-roll solution presented in sec. \ref{esr} and imposes BDic at 
$ \tau_0 = -\infty $ (that is, $ \eta_0 = -\infty $) then $ D(k) = 0 $ and the fluctuation 
power spectrum at the end of inflation is the usual power spectrum
$ P_\mathcal{R}(k)=P^{BD}_{\mathcal{R}}(k) $.

\medskip

\begin{figure}[h]
\includegraphics[width=16.cm]{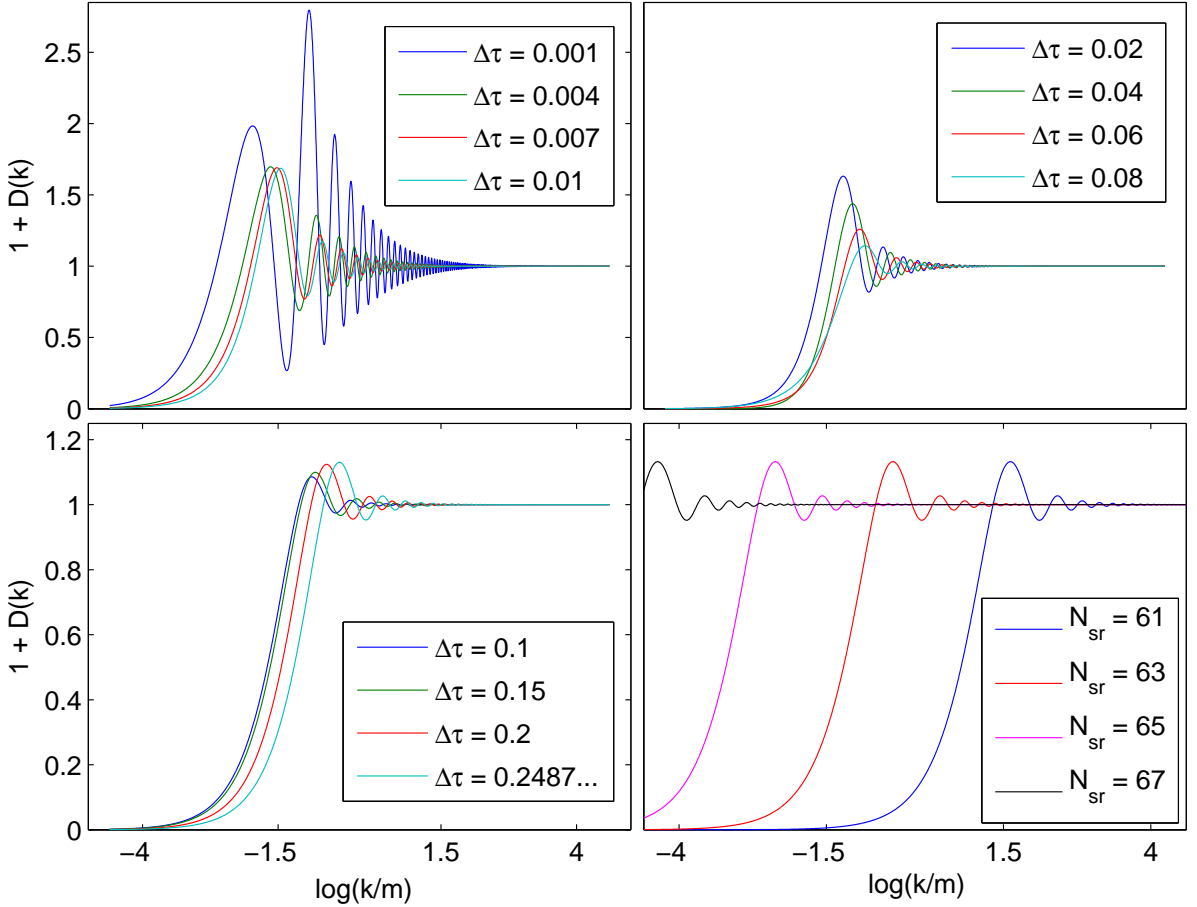}
\caption{Numerical transfer function $ 1 + D(k) $.
Lower left panel: Numerical transfer function $ 1 + D(k) $ for
BDic at $ \tau = \tau_0 = \tau_\ast + \Delta \tau $, for different $ \Delta \tau $ 
values as given in the picture. We see here that the peak of $ 1 + D(k) $ grows and moves for
larger $ k $ as $ \tau_0 $ increases. Here $ N_{sr} = 63 $.
Lower right panel: The transfer function $ 1+D(k) $ when the BDic eq.(\ref{BD}) are imposed
during slow--roll at finite times $ \tau_0 $ and $ N_{sr} $ efolds of slow--roll 
have still to occur.
Upper panels: Numerical transfer function $ 1 + D(k) $ for
BDic at $ \tau = \tau_0 = \tau_\ast + \Delta \tau $, for different values of 
$ \Delta \tau $ as given in the picture. We get stronger oscillations in  
$ 1 + D(k) $ for decreasing $ \tau_0 $ in
the range $ \Delta \tau < 0.04 $. Here $ N_{sr} = 63 $.}
\label{fig:DBDC}
\end{figure}

$ P^{BD}_{\mathcal{R}}(k) $ is given by its customary slow-roll expression,
\be\label{pbd}
 \log  P^{BD}_{\mathcal{R}}(k) = \log A_s(k_0) + (n_s-1) \; \log\frac{k}{k_0} + 
\tfrac12 \; n_{run} \;  \log^2\frac{k}{k_0} +{\cal O}\left(\frac1{N^3}\right) \; .
\ee
We solved numerically the fluctuations equation (\ref{flutau}) in
cosmic time with the BDic eq.(\ref{BD}) covering both the fast-roll and slow-roll
regimes. We started at
initial times $ \tau_0 $ ranging from the vicinity of 
$ \tau = \tau_* $ till the transition
time $ \tau_{trans} = 0.2487963\ldots $ from fast-roll to slow-roll.
We  computed the transfer function $ D(k) $ from the mode functions
behaviour deep during slow-roll inflation from eqs.(\ref{curvapot}) and (\ref{powR})
\cite{biblia}. 
In figs. \ref{fig:DBDC} we depict $ 1 + D(k) $ vs. 
$ k $ for twelve values of the time $ \tau_0 $ where BDic are imposed.

\medskip

\begin{figure}[h]
\includegraphics[width=14.cm]{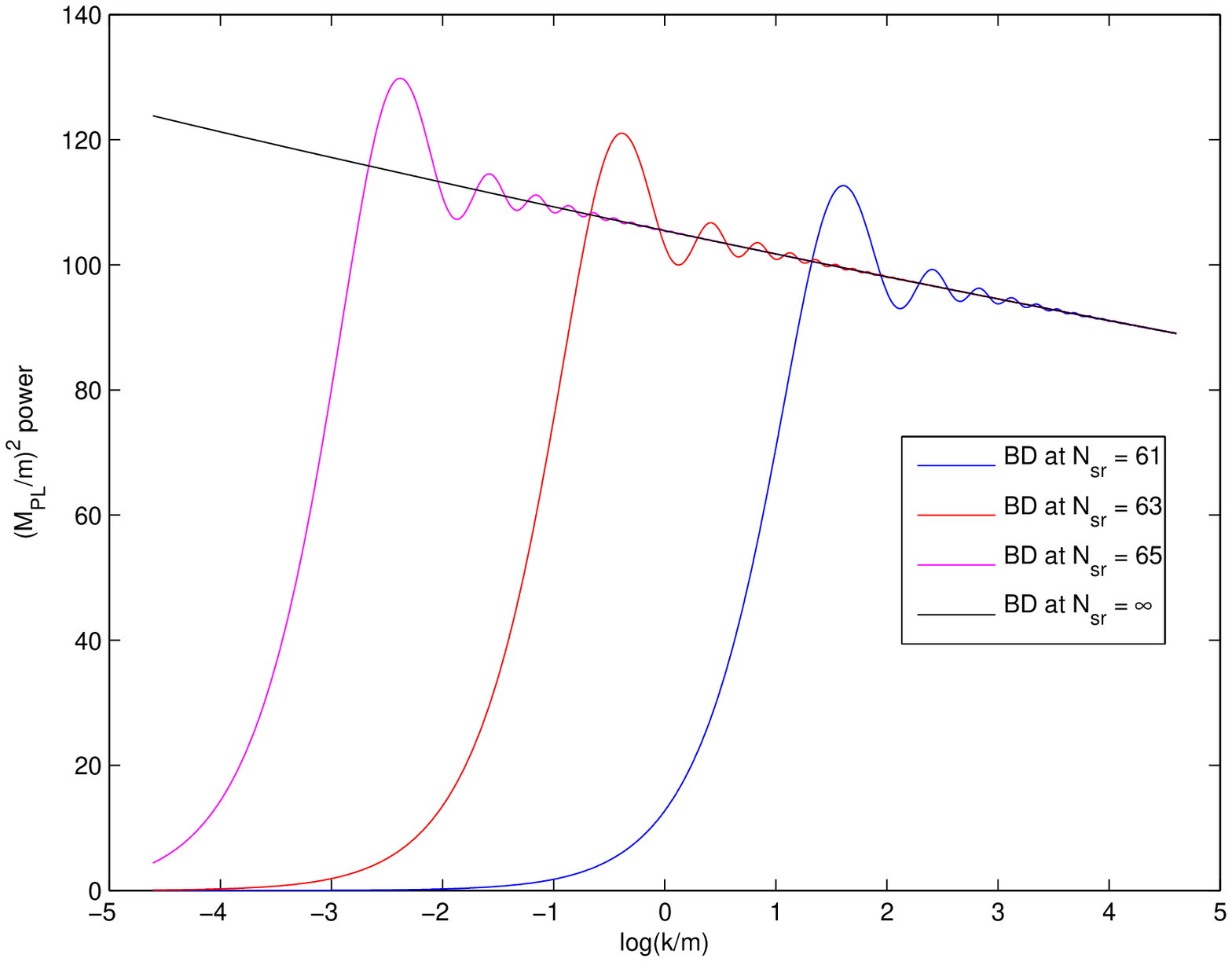}
\caption{Power spectrum with BDic eq.(\ref{BD}) imposed during slow--roll
when $ N_{sr} $ efolds of slow--roll inflation have still to occur. We see here the decrease of
the power spectrum $ P_\mathcal{R} $ as $ k^{n_s-1} $ multiplied by the oscillations
of $ 1 + D(k) $. See eqs.(\ref{powR}) and (\ref{pbd}) and figs. \ref{fig:DBDC}.
The non-oscillatory black curve corresponds to the usual power with BDic at $ \eta_0 =
-\infty $ eq.(\ref{pbd}) decreasing as $ k^{n_s-1} $. The later are imposed the BDic, 
the smaller is the number of slow-roll efolds $ N_{sr} $ and the whole $k$-spectrum 
 shifts to larger $ k $.}
\label{fig:BDs}
\end{figure}

Notice that when BDic are imposed at {\bf finite times} $ \tau_0 $, the spectrum 
{\bf is not} the usual $ P^{BD}_{\mathcal{R}}(k) $ but it gets modified by 
a non-zero transfer function $ D(k) $ 
eq.(\ref{powR}). The power spectrum $ P_{\mathcal{R}}(k) $ vanishes at $ k = 0 $  
and exhibits oscillations which vanish at large $ k $ [see figs. \ref{fig:DBDC}
and \ref{fig:BDs}].

\medskip

During slow-roll different initial times $ \tau_0 $ lead essentially to a rescaling of $ k $
in $ D(k) $ by a factor  $ \eta_0 $ since the 
conformal time $ \eta $ is almost proportional to $ 1/a(\eta) $ during slow-roll 
[see figs.~\ref{fig:BDs}- \ref{fig:DBDC} and below eq.(\ref{dsr})]. 
By virtue of the dynamical attractor character of slow--roll, 
the power spectrum when the BDic are imposed at a finite time 
$ \tau_0 $ cannot really distinguish between the extreme slow--roll solution 
(for which slow--roll starts from the very beginning $ \eta_0 = -\infty $)
or any other solution which is attracted to slow--roll well before the time $ \tau_0 $.

\subsection{Accurate numerical computation of the power spectrum and the
 transfer function $ D(k) $ of initial conditions.}

In order to accurately calculate $ n_s $ we proceed as follow. 
We match the solution $ S_\mathcal{R}(k;\eta) $ with the slow--roll 
solution  $ g_{\nu_\mathcal{R}}(k;\eta) $ eq.(\ref{gnu})  at the time $ \tau_0 $ when 
$ N_{sr} $ efolds
of slow--roll have still to occur.  $ \eta $ and $ \nu_\mathcal{R} $ are computed at this time 
$ \tau_0 $. In practice, this corresponds to setting 
$ A_\mathcal{R}(k)=1, \; B_\mathcal{R}(k) = 0 $ 
(and therefore $ D_\mathcal{R}(k) = 0 $) in the Bogoliubov transformation eq.(\ref{bogo}).  

Then, we integrate numerically the fluctuations
equations  eq.(\ref{flutau}). By construction, this produces the standard spectra 
$ P^{BD}_{\mathcal{R}}(k) $ eq.(\ref{pbd}) that quickly stabilize as $ N_{sr} $ is 
increased a few efolds above $ N=60 $.

\medskip

It is convenient to introduce the quantity
\be\label{defl}
L_s \equiv \log\left[\left(\frac{M_{PL}}{m}\right)^2 \; A_s(k_0=m)\right] \; ,
\ee
with $ k_0=m $ when $ a(\eta)=1 $, that is $ N=60 $ efolds before inflation ends. 
In table IV we provide $ L_s $ for several values of $ N_{sr} $. 

\begin{table}
\begin{tabular}{|c|c|c|c|}
\hline  
   ~~$N_{sr}$ ~~&~~ $L_s$ ~~&~~ $n_s$ ~~&~~ $ n_{run} $  \\
\hline \hline 
   ~~61~~&~~ $4.6585381\ldots$ ~~&~~ $0.9637013\ldots$ ~~&~~ $-0.0000701\ldots$ \\
\hline
   ~~63~~&~~ $4.6583004\ldots$ ~~&~~ $0.9641135\ldots$ ~~&~~ $-0.0001639\ldots$ \\
\hline
   ~~65~~&~~ $4.6584371\ldots$ ~~&~~ $0.9642483\ldots$ ~~&~~ $-0.0002165\ldots$ \\
\hline
   ~~67~~&~~ $4.6584463\ldots$ ~~&~~ $0.9642444\ldots$ ~~&~~ $-0.0002165\ldots$ \\
\hline 
   ~~69~~&~~ $4.6584469\ldots$ ~~&~~ $0.9642448\ldots$ ~~&~~ $-0.0002167\ldots$\\
\hline
\end{tabular}
\caption{Exact values of $ L_s = \log[\left(M_{PL}/m\right)^2 \; A_s(k_0=m)] $
for several values of $ N_{sr} $ from the numerical calculation. The exact values of
$ n_s $ vary little with $ N_{sr} $ and are close to the slow-roll approximation value. 
Also $ n_{run} $ is close to the value in the slow-roll approximation.}
\end{table}

\medskip

To transform this $ k_0 $ in a wavenumber today we need:  
\begin{itemize}
\item {the total redshift from 60 efolds before inflation ends till today
[since we choose $ a( \tau = 0 ) = 1 $  when there are still $ N=60 $ 
efolds till the end of inflation].}
\item {the value of $ m $ as determined by the observed value of the amplitude $ A_s(k_0) $.}
\end{itemize}

Let $ k_0^{\rm CMC} $ be the
value of the pivot scale of CosmoMC [that is 50 (Gpc)$^{-1}$ today] 60 efolds
before the end of inflation. Then, we have from eqs.(\ref{pbd}) and (\ref{defl}), 
\be\label{Asm}
\log A_s(k_0=m) = L_s + 2 \, \log\frac{m}{M_{PL}} =  L_s^{\rm CMC} +
  (n_s^{\rm CMC}-1) \; \log\frac{m}{k_0^{\rm CMC}} + \tfrac12  \; n_{run} \; 
\left[\log\frac{m}{k_0^{\rm CMC}}\right]^2 + {\cal O}\left(\frac1{N^3}\right) \; ,
\ee
where $ L_s^{\rm CMC} \equiv \log A_s^{\rm CMC} (k_0^{\rm CMC}) $ and $ n_s^{\rm CMC} $ 
are best fit values in a given CosmoMC run. Since the running index $ n_{run} $ is 
$ {\cal O}\left(1/N^2\right) $, we get for $ m $,
\be
  \left(\frac{m}{M_{PL}}\right)^2 = \left(\frac{m}{k_0^{\rm CMC}}\right)^{n_s^{\rm CMC}-1}
  \exp\,(L_s^{\rm CMC} - L_s)\left[1+  {\cal O}\left(\frac1{N^2}\right)\right] \; .
\ee
The wavevectors at $ a = 1 $ (60 efolds before inflation ends) and today are related by
\cite{biblia}
\be\label{ka1}
k^{a=1} = \frac{e^{60}}{a_r} \;  k^{today} \; ,
\ee
where $ a_r $ is the scale factor by the end of inflation
\be\label{var}
a_r = 2.5 \; 10^{-29} \; \sqrt{\frac{10^{-4}M_{PL}}{H_{60}}} \; ,
\ee
and $ H_{60} $ is the Hubble parameter 60 efolds before inflation ends.
We thus have for the pivot wavenumber at $ a = 1 $
\be \label{vk0}
k_0^{\rm CMC} \simeq 1.46\ldots \, \sqrt{\frac{H_{60}}{10^{-4}M_{PL}}} \times 10^{15}~{\rm GeV}
\ee
and
\begin{equation*}
  \left(\frac{m}{M_{PL}}\right)^{2-(n_s^{\rm CMC}-1)/2} = 
   \left(\frac{16.67\ldots}{\sqrt{h_{60}}}\right)^{n_s^{\rm CMC}-1} \,
   \exp\,(L_s^{\rm CMC} - L_s) \quad , \quad
{\rm where} \quad h_{60} \equiv \frac{H_{60}}{m} \; .
\end{equation*}
Notice the small $ 1/N $ correction $ (n_s^{\rm CMC}-1)/2 $ in the exponent of
$ m/M_{PL} $. Eq.(\ref{Asm}) yields for the best fit CosmoMC run 
$ L_s^{\rm CMC} = -19.9808\ldots $ and $  n_s = 0.9635\ldots $ \cite{biblia}:
\begin{equation*}
   m \simeq 4.8114\ldots 10^{-6}\, M_{PL} =
         1.1717 \ldots 10^{13}~{\rm GeV}
\end{equation*}
The exact values given above in Table IV
$$ 
A_s =  \left(\frac{m}{M_{PL}}\right)^2 \; \exp(L_s) \; ,  \quad n_s   \quad 
{\rm and}  \quad  n_{run} 
$$
are obtained taking into account 
the fast-roll and slow--roll stages in the numerical calculation.
We can compare them to their slow--roll (leading $ 1/N $) analytic
counterparts for the double-well quadratic plus quartic potential,
\cite{biblia}
\begin{equation*}
  A_s = \frac{N^2}{12\pi^2} \left(\frac{m}{M_{PL}}\right)^2 \, 
  \frac{(1-z)^4}{y^2 \, z} \;,\quad 
  n_s = 1 - \frac{y}{N} \, \frac{3 \, z + 1}{(1-z)^2} \;,\quad 
  n_{run} =  \frac{y^2 \, z}{N^2 \, (1-z)^4}\left(24 \, z^2 - 35 \, z +3\right)
\end{equation*}
where $ N=60 , \; z = 0.117446 $ and $ y = z-1-\log z = 1.2592226\ldots $, that is
\begin{equation*}
  A_s = \frac{N^2}{12\pi^2} \; \left(\frac{m}{M_{PL}}\right)^2 \; 
\exp(4.59536898\ldots) \; , \quad   n_s = 0.9635620\ldots \; , \quad 
  n_{run} = -0.0000664 \ldots  
\end{equation*}
The figure in the exponent is to be compared with the $ L_s $ values in Table IV.
The agreement with Table IV is quite good, especially for $ n_s $.

\medskip

\begin{figure}[h]
\includegraphics[width=16.cm]{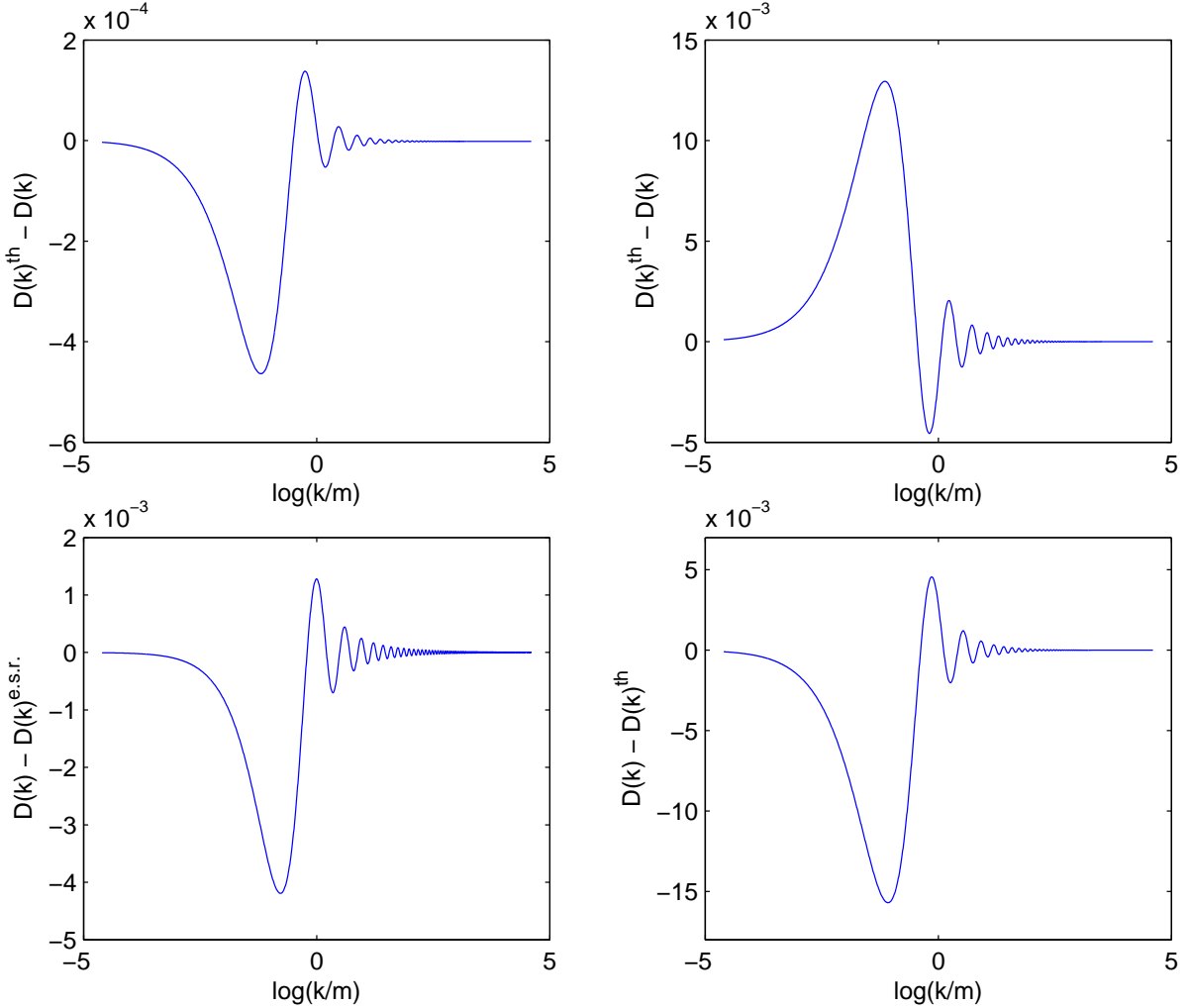}
\caption{Upper Left panel: Difference between the (approximate) transfer function 
$ {\tilde D}(k \; \eta_0) $ eqs.(\ref{dsr})-(\ref{dsr32}) for 
$ \nu_\mathcal{R}=1.5182189\ldots $ and the numerical 
(exact at least to a $ 10^{-7} $ relative error) transfer function 
$ D(k) $, when $ N_{sr} = 63 $. Upper Right  panel: Difference between the 
(approximate) transfer function $ {\tilde D}(k \; \eta_0) $ eqs.(\ref{dsr})-(\ref{dsr32}) for
$ \nu_\mathcal{R}=3/2 $ (the scale-invariant value) and the numerical (exact)
transfer function. We see that the difference in the right panel [eq.(\ref{dsr32})] is 
$ < 0.014 $ while in the left panel the difference of the analytic formula eq.(\ref{dsr}) 
is much smaller, $ < 0.0005 $.
Lower Left panel: difference between the exact (numerical) $ D(k) $ computed for 
the fast-roll inflaton solution of table II and for the extreme slow--roll inflaton solution
of table I when BDic are imposed 63 efolds before the end of inflation. Lower Right 
panel: difference between the numerical (exact) fast--roll $ D(k) $ and the approximate 
$ {\tilde D}(k \; \eta_0) $ calculated with $ \nu_\mathcal{R}=3/2 $ and 
$ \eta_0=-4.0169827\ldots $. We see that the differences are small in both cases.}
\label{fig:deltaD}
\end{figure} 

We now find the exact (numerical) transfer 
function $ D(k) $ for the initial conditions, by simply taking in eq.(\ref{powR})
the ratio of the two power spectra: $ P_{\mathcal{R}}(k) $ with BDic at time $ \tau_0 $ 
and  $ P^{BD}_{\mathcal{R}}(k) $. 
In the case of BDic at finite times the result is given in fig~\ref{fig:DBDC}.
At the largest value $ k/m=100 $ of the wavenumber interval considered, we have
$$
 1+D(100 \, m)= 0.9996994\ldots , \; 1.0000061\ldots, \; 1.0000001\ldots 
$$ 
for 
$$
N_{sr}=61, \; 63, \; 65  \; {\rm and} \;   67 ,  \; {\rm respectively}. 
$$
This provides a good check of the accuracy of the calculation.

In figs. \ref{fig:deltaD} we compare the numerically computed $ D(k) $ 
against $ {\tilde D}(k \; \eta_0) $ analytically computed for BDic imposed at time $ \eta_0 $
during slow--roll in eq.(\ref{dsr}), sec. \ref{bdicsr}.
The comparison is performed for BDic imposed when $ N_{sr} = 63 $ on the
extreme slow roll solution, which corresponds to $ \eta_0=-4.0202308\ldots $. We
consider two values of $ \nu_\mathcal{R} : \;  \nu_\mathcal{R} = 2-n_s/2 = 1.5182189\ldots $,
$ n_s=0.9635620\ldots $ corresponding to slow--roll at leading $ 1/N $ order, and the
exactly scale-invariant case $ \nu_\mathcal{R}=3/2 $. 
Notice that in the latter case 
$ {\tilde D}(k \; \eta_0) $ has the explicit simple analytic form eq.(\ref{dsr32}).

The maximum of the numerical transfer function $ 1+D(k) $ is located at $ k/m=0.68755\ldots$ 
and has the value $ 1.13218\ldots $ The maximum of $ 1+ {\tilde D}(k \; \eta_0) $, when 
$ \nu_\mathcal{R}=3/2 $ is in $ k/m=0.68755\ldots $ and has the value 
$ 1.13009\ldots $. Recall that these values of $ k/m $ have the scale fixed 
by the choice $ a=1 $ when $ N=60 $ efolds lack
before inflation ends. 

\medskip

Let us now consider the fluctuations on the fast--roll solution of 
Table II. Since $ \eta $ has a finite lower limit, the choice 
$ A_\mathcal{R}(k)=1, \; B_\mathcal{R}(k) = 0 $ has little 
meaning and BDic can be imposed only at a {\rm finite} time $ \tau_0 $ later than 
the singularity time $ \tau_\ast $. If $ \tau_0 $ is exactly the transition time 
$ \tau_{trans} $ when $ \epv=1/N $, fast-roll ends and slow--roll begins, 
(to proceed for $ N_{sr}=63 $ efolds), then $ D(k) $ does not differ too much 
from that computed with the extreme slow roll solution. 
This comparison is performed in the lower left 
panel of fig.~\ref{fig:deltaD}. In the right panel $ D(k) $ is compared
to the $  {\tilde D}(k \; \eta_0) $ for $ \nu_\mathcal{R}=3/2 $ and $ \eta_0=-4.0169827\ldots $, 
which is the value of the conformal time at the onset of slow--roll 
(see Table II). 

\medskip

When the BDic are imposed during the fast--roll stage well {\bf before} it ends, 
$ D(k) $ changes much more significantly than along the extreme slow roll solution. 
This is due to two main effects: the
potential felt by the fluctuations is attractive during fast--roll and $ \eta_0 $, far from 
being almost proportional to $ 1/a(\eta) $, tend to the constant value 
$ \eta_\ast $ as $ \tau \to \tau_\ast^+ $ and $ a(\eta)\to 0 $. The numerical 
transfer functions $ 1+ D(k) $ obtained from eqs.(\ref{curvapot}) and (\ref{powR})
are plotted in figs.~\ref{fig:DBDC}.
 
\medskip

The fact that choosing BDic leads to a primordial power and its respective
CMB multipoles which correctly {\bf reproduce} the observed spectrum 
justifies the use of BDic for the scalar curvature fluctuations.

\subsection{The effect of the fast-roll stage on the low multipoles of the 
CMB}\label{eiclm}

In the region of the Sachs-Wolfe plateau for $ l \lesssim 30 $, the
matter-radiation transfer function can be set equal to unity and the
CMB multipole coefficients $ C_l's $ are given by \cite{hu} 
\be \label{cl}
C_l =\frac{4\pi}{9} \int_0^\infty \frac{dk}{k}\,  {P}_X(k) \left\{
j_l[k(\eta_0-\eta_{LSS})]\right\}^2  \; , 
\ee 
where $ P_X $ is the power spectrum of the corresponding perturbation, $
X=\mathcal{R} $ for curvature perturbations and $ X=T $ for tensor
perturbations,  $ j_l(x) $ are spherical Bessel functions \cite{abra} 
and $ \eta_0-\eta_{LSS} $ is the comoving
distance between today and the last scattering surface (LSS)
given by 
\be \label{etaLSS}
\eta_0-\eta_{LSS} = \frac1{H_0} \; \int^1_{\frac1{1+z_{LSS}}}
\frac{da}{\sqrt{\Omega_r + \Omega_M \; a + \Omega_{\Lambda} \; a^4}} \; , 
\ee 
where $ \Omega_r , \; \Omega_M $ and $ \Omega_{\Lambda} $ stand for the 
fraction of radiation, matter and cosmological constant in today's 
Universe. We find using $ z_{LSS} =1100 $,
\be\label{329}
\eta_0-\eta_{LSS} = \frac{3.296}{H_0}  \; . 
\ee 
Notice that $ k/H_0 \sim d_H/\lambda_{phys}(t_0) $ is the ratio
between today's Hubble radius and the physical wavelength. The power
spectrum for curvature ($ \mathcal R $) perturbations 
$ P_{\mathcal{R}}(k) $ is given by eqs.(\ref{powR})-(\ref{pbd}). 

\begin{figure}[h]
\includegraphics[width=16.cm]{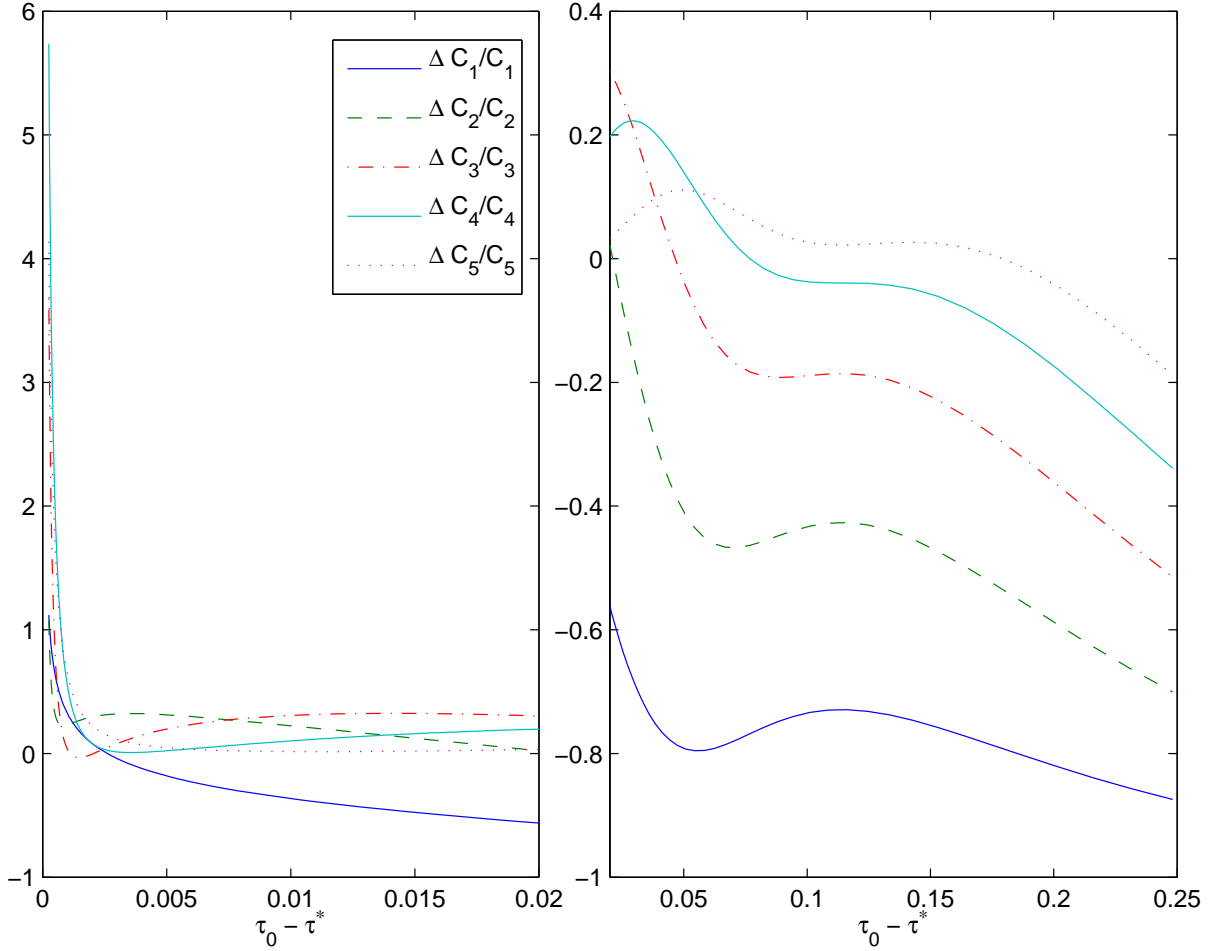}
\caption{The change $ \Delta C_{\ell}/ C_{\ell} $ on the 
CMB multipoles for $ \ell =1, \ldots, 5 $.
Upper plot:  $ \Delta C_{\ell}/ C_{\ell} $ vs. $ \tau_0 - \tau_* $ for $ 0<\tau_0 - \tau_* < 
0.2487963\ldots$. Lower plot:  $ \Delta C_{\ell}/ C_{\ell} $ vs. $ \tau_0 - \tau_* $ for 
$ 0.0193<\tau_0 - \tau_* < 0.2487963\ldots . \; \tau_0 $ is the time when the BDic 
eq.(\ref{BD}) are imposed to the fluctuations. We choose $ \tau_0 $ 
inside the fast-roll stage. $ \Delta C_{\ell}/ C_{\ell} $ is {\bf positive} for 
small $ \tau_0  - \tau_\ast $ and {\bf decreases} with $ \tau_0 $ becoming then 
{\bf negative}. The CMB quadrupole observations indicate a large {\bf suppression}
thus indicating that $ \tau_0 - \tau_\ast \gtrsim 0.05 \simeq 10100 \; \tau_{Planck} $.
Our {\bf predictions} here for the quadrupole and octupole suppressions 
are to be confronted with forthcoming CMB observations.
It will be extremely interesting to measure the primordial dipole and compare
with our predicted value.} 
\label{dcl}
\end{figure}

Inserting eq.(\ref{powR}) into  eq.(\ref{cl}) yields the $ C_l $ as the sum
of two terms
\be \label{DelC}
C_l = C ^{BD}_l + \Delta C_l
\quad , \quad \frac{\Delta C_l}{C_l} = \frac{\int^\infty_0
D(\kappa \; x)~ f_l(x) \; dx}{\int^\infty_0 f_l(x) \; dx} \quad , \quad
x= k(\eta_0-\eta_{LSS}) = k/\kappa \; ,
\ee 
where from eq.(\ref{329}), $ \kappa \equiv H_0/3.296\ldots $, 
\be \label{fls} 
f_l(x)= x^{n_s-2} \; [j_l(x)]^2 \; . 
\ee
and $ j_l(x) $ stand for the spherical Bessel functions.

The $ C ^{BD}_l $'s correspond to the standard BD power spectrum  
$ P^{BD}_{\mathcal{R}}(k) $ eq.(\ref{pbd}) and the $ \Delta C_l $ exhibit 
the effect of the transfer function $ D(k) $ on the $ C_l $.

\medskip

Using the transfer function $ D(k) $ obtained above eq.(\ref{DofkR}), 
we computed the change on the 
CMB multipoles $ \Delta C_{\ell}/ C_{\ell} $ for $ \ell =1,\ldots, 5 $  as
functions of the starting instant of the fluctuations $ \tau_0 $.
We plot $ \Delta C_{\ell}/ C_{\ell} $ for $ 1 \leq \ell \leq 5 $ vs. $ \tau_0 - \tau_\ast $
in fig. \ref{dcl}. We see that $ \Delta C_{\ell}/ C_{\ell} $ is {\bf positive} for 
small $ \tau_0  - \tau_\ast $ and {\bf decreases} with $ \tau_0 $ becoming then 
{\bf negative}. The CMB quadrupole observations indicate a large {\bf suppression}
thus indicating that $ \tau_0 - \tau_\ast \gtrsim 0.05 \simeq 10100 \; \tau_{Planck} $.

Being $ D(k) < 0 $ for low $ k $ as depicted in figs. \ref{fig:DBDC},
the primordial power at large scales is then suppresed and
the low $ C_{\ell} $ decrease as seen from eq.(\ref{DelC}).

$ \Delta C_{\ell}/ C_{\ell} $ mainly originates from the peak of $ D(k) $ displayed in figs.
 \ref{fig:DBDC} whose position moves to smaller $ k $ for 
decreasing $ \tau_0 $. Therefore, the primordial power suppression is
less important for decreasing $ \tau_0 $ and the CMB multipole suppression
$ \Delta C_{\ell}/ C_{\ell} $ less important as depicted in figs. \ref{dcl}.

For small $ \tau_0 - \tau_\ast \lesssim 0.05 $ the peak of $ D(k) $ grows
significantly and $ \Delta C_{\ell}/ C_{\ell} $ become positive, namely the low CMB multipoles
are enhanced.

\medskip

It should be recalled that the observation of a low CMB quadrupole 
sparked many different proposals to explanain that suppression \cite{expla}.
\medskip

Besides finding a CMB quadrupole suppression in agreement with observations 
\cite{biblia}-\cite{quamc},
we provide here {\bf predictions} for the dipole and octupole suppressions.
Forthcoming CMB observations can provide better data to confront our
quadrupole and octupole suppression predictions.
It will be extremely interesting to measure the primordial dipole and compare
with our predicted value.

\section{Analytic formulas for the transfer function $ D(k) $.}

It is very important to dispose of analytic formulas for the transfer 
function $ D(k) $ in order to better understand the physical origin
of its oscillations and properties as well as in 
the perspective of the MCMC data analysis.

However, the mode equations (\ref{fluces}) are not solvable in closed 
form for $ k \neq 0 $ , not even for the approximated inflation solution 
eq.(\ref{solfr}) which leads to the potential $ V_\mathcal{R}(\tau) $ 
eq.(\ref{vafr}).

The function $ D(k) $ must obey the general properties eq.(\ref{propg}).

\subsection{The primordial power spectrum vanishes for $ k \to 0 $
and becomes the BD power spectrum  for $ k \to \infty $}\label{dcero}

The fluctuations equation (\ref{fluces}) can be solved explicitly for $ k = 0 $
\be\label{S0}
s(\eta) = c_1 \; z(\eta) + c_2 \; z(\eta) \; 
\int_{\eta_0}^{\eta} \frac{d\eta'}{z^2(\eta')} \; ,
\ee
where $ c_1 $ and $ c_2 $ are arbitrary constants.

The BDic eq.(\ref{BD}) introduce for $ k \to 0 $ a $ 1/\sqrt{2 \, k} $ singularity
in the mode functions. Thus, the mode functions must have the behaviour
\be\label{srk0}
 S_\mathcal{R}(k;\eta)\buildrel{k \to 0 }\over= \frac{s(\eta)}{\sqrt{2 \, k}}\left[ 1 +
{\cal O}(k) \right] 
\ee
where $ s(\eta) $ is given by eq.(\ref{S0}).

Inserting eq.(\ref{srk0}) into the BDic eq.(\ref{BD}) yields for $ k \to 0 $,
$$
s(\eta_0) = 1 \quad , \quad \frac{ds(\eta_0)}{d\eta} = 0 \; ,
$$
which determines the coefficients $ c_1 $ and $ c_2 $ in eq.(\ref{S0}).
We finally obtain
\be\label{sfin0}
s(\eta) = \frac{z(\eta)}{z(\eta_0)}- z'(\eta_0) \; z(\eta) \; 
\int_{\eta_0}^{\eta} \frac{d\eta'}{z^2(\eta')}
\ee
and using eq.(\ref{zeta2}) valid for $ \eta \to  0^- $ when slow--roll applies
\be\label{sfin1}
{\displaystyle \lim_{\eta \to  0^-}}\frac{s(\eta)}{z(\eta)} = \frac1{z(\eta_0)} \; .
\ee

The primordial power spectrum for $ k \to 0 $ follows by inserting 
eq.(\ref{srk0}) and eq.(\ref{sfin1}) into the general expression eq.(\ref{curvapot}), 
$$
P_\mathcal{R}(k) \buildrel{k \to 0 }\over= \left(\frac{m}{M_{PL}}\right)^2
\; \frac{k^3}{2 \; \pi^2} \; {\displaystyle \lim_{\eta \to  0^-}}
\left|\frac{S_\mathcal{R}(k;\eta)}{z(\eta)} \right|^2 \buildrel{k \to 0 }\over=
\left(\frac{m}{M_{PL}}\right)^2 \; \left( \frac{k}{2 \, \pi \; z(\eta_0)}\right)^2
$$
We thus find in general that the power spectrum vanishes as $ k^2 $ for $ k \to 0 $ 
and therefore 
$$
1 + D(k) \buildrel{k \to 0 }\over= {\cal O}(k^{n_s+1})
$$
as stated in eq.(\ref{propg}). This property is generally true except for the extreme
slow-roll inflaton solution (sec. \ref{esr}) with BDic imposed at $ \eta_0 = -\infty $
in which case $ D(k) $ vanishes identically for all $ k $.

\bigskip

For growing $ k $ the modes exit the horizon later on, 
during the slow--roll regime where eq.(\ref{bogo}) applies. For large $ k $
the mode functions $ S_\mathcal{R} $ as well as $ g_{\nu_\mathcal{R}} $ 
behave as plane waves [eqs.(\ref{BDk}) and (\ref{geta0})] and therefore  
$$
A_\mathcal{R}(k)=1 \quad , \quad B_\mathcal{R}(k) = 0 \quad .  \quad
{\rm Hence} \quad D(k) \buildrel{k \to \infty }\over= 0 \; .
$$.

\subsection{The transfer function $ D(k) $ when BDic are imposed during 
slow--roll.}\label{bdicsr}

When the BDic eq.(\ref{BD}) are imposed during slow--roll at a finite time $ \eta_0 $
we can use eq.(\ref{bogo}) for the mode functions at $ \eta = \eta_0 $ and we obtain,
\bea
\frac{e^{-i \; k \; \eta_0}}{\sqrt{2 \; k}} &=&
A_\mathcal{R}(k) \;
g_{\nu_\mathcal{R}}(k;\eta_0) + B_\mathcal{R}(k) \; g^*_{\nu_\mathcal{R}}(k;\eta_0) \cr \cr
-i \; k \; \frac{e^{-i \; k \; \eta_0}}{\sqrt{2 \; k}} &=&
A_\mathcal{R}(k) \;
g'_{\nu_\mathcal{R}}(k;\eta_0) + B_\mathcal{R}(k) \; g'^*_{\nu_\mathcal{R}}(k;\eta_0) 
\eea
which determines
\be\label{absr}
A_\mathcal{R}(k)=\frac{e^{-i \; k \; \eta_0}}{i \; \sqrt{2 \; k}}\left[ 
g'^*_{\nu_\mathcal{R}}(k;\eta_0) + i \; k \;  g^*_{\nu_\mathcal{R}}(k;\eta_0) \right]
\quad ,  \quad
B_\mathcal{R}(k)=\frac{e^{-i \; k \; \eta_0}}{i \; \sqrt{2 \; k}}\left[ 
g_{\nu_\mathcal{R}}'(k;\eta_0) + i \; k \;  g_{\nu_\mathcal{R}}(k;\eta_0) \right] \; .
\ee
These coefficients satisfy eq.(\ref{abc}) and
$$
|A_\mathcal{R}(k)|^2 + |B_\mathcal{R}(k)|^2 = \frac1{k}\left[ 
|g_{\nu_\mathcal{R}}'(k;\eta_0)|^2 + k^2 \;  |g_{\nu_\mathcal{R}}(k;\eta_0)|^2 \right]
$$
Notice that the function $ g_{\nu}(k;\eta) $ eq.(\ref{gnu}) and the $ k $ factors
in eq.(\ref{absr}) combine to produce functions $ A_\mathcal{R}(k) \equiv {
\tilde A}_\mathcal{R}(k \; \eta_0) $ and $ B_\mathcal{R}(k) \equiv
{\tilde B}_\mathcal{R}(k \; \eta_0) $ that only depend on the product $ k \; \eta_0 $. 

We find from eqs.(\ref{DofkR}) and (\ref{absr}) the corresponding
transfer function which is a function of $ k \; \eta_0 $ too,
\be\label{dsr}
1+{\tilde D}(k \; \eta_0) = \frac1{k} \left\{ |g_{\nu_\mathcal{R}}'(k;\eta_0)|^2 + 
k^2 \;  |g_{\nu_\mathcal{R}}(k;\eta_0)|^2- {\rm Re} \left[i^{3-2 \, \nu_\mathcal{R}}
\; \left( g'^2_{\nu_\mathcal{R}}(k;\eta_0) + k^2 \;  g^2_{\nu_\mathcal{R}}(k;\eta_0)
\right)\right]\right\} 
\ee
The functional dependence on $ k \; \eta_0 $ confirms the assertion in sec. \ref{priflu} 
that different initial times $ \tau_0 $ lead to a rescaling in $ k $.

\medskip

In the $ k \; \eta_0 \to \infty $ limit two types of vanishing terms show up in 
$ {\tilde D}(k \; \eta_0) $: (a) terms that strongly oscillate as 
$ e^{\pm 2 \, i \; k \; \eta_0} $ as they tend to zero and (b) non-oscillatory
decreasing terms. Under integrals on $ k $, the terms of type (a) yield convergent expressions.
We derive the non-oscillatory decreasing terms (b) by inserting the asymptotic behaviour of the
Hankel functions eq.(\ref{gnu}) \cite{abra} in eq.(\ref{dsr}) with the result
\be \label{dkasi}
{\tilde D}(k \; \eta_0)  \buildrel{k \to \infty}\over= 
\frac{(\nu^2 - \frac14)^2}{8 \; (k \; \eta_0)^4 } + 
{\rm terms ~ oscillating ~ as}~  e^{\pm 2 \, i \; k \; \eta_0} \; .
\ee
However, this approximation will not be valid for large enough $ k $ since the
modes at small enough wavelength will exit the horizon after the end of slow--roll
where eq.(\ref{dkasi}) does not apply anymore. We recall that 
the occupation number $ |B_\mathcal{R}(k)|^2 $ (and therefore $ D(k) $) must 
decrease faster than $ 1/k^4 $ for
$ k \to \infty $ in order to ensure finite UV values for the expectation 
value of the energy-momentum fluctuations \cite{biblia,motto2}. 

\medskip

The case $ \nu_\mathcal{R} = 3/2 $ is a good approximation 
which simplifies the expressions above. We obtain in this scale invariant case:
$$
A_\mathcal{R}(k) = 1 + \frac{i}{k \; \eta_0} - \frac1{2 \; k^2 \; \eta_0^2}
\quad , \quad B_\mathcal{R}(k) =- \frac{e^{-2 \, i \; k \; \eta_0}}{2 \; k^2 \; \eta_0^2}
\; .
$$
The transfer function is in this case,
\be\label{dsr32}
{\tilde D}(x) = \frac{\cos 2x}{x^2} - \frac{\sin 2x}{x^3}  + \frac{\sin^2 x}{x^4}
\quad , \quad  \nu_\mathcal{R} = 3/2 \quad , \quad x \equiv k \; \eta_0 \; .
\ee
Eq.(\ref{dkasi}) for $ \nu = 3/2 $ coincides with eq.(\ref{dsr32}) in the 
 $ x \to \infty $ limit, as it must be.

Notice that the simple formula eq.(\ref{dsr32}) obeys the general properties eq.(\ref{propg}).
In particular,
$$
{\tilde D}(x) \buildrel{x \to 0 }\over= -1 + \frac49 \; x^2 + {\cal O}(x^4) \; .
$$

\section{Fixing the Total Number of Inflation e-folds and the bound 
from Entropy}

\begin{figure}[h]
\includegraphics[width=16.cm]{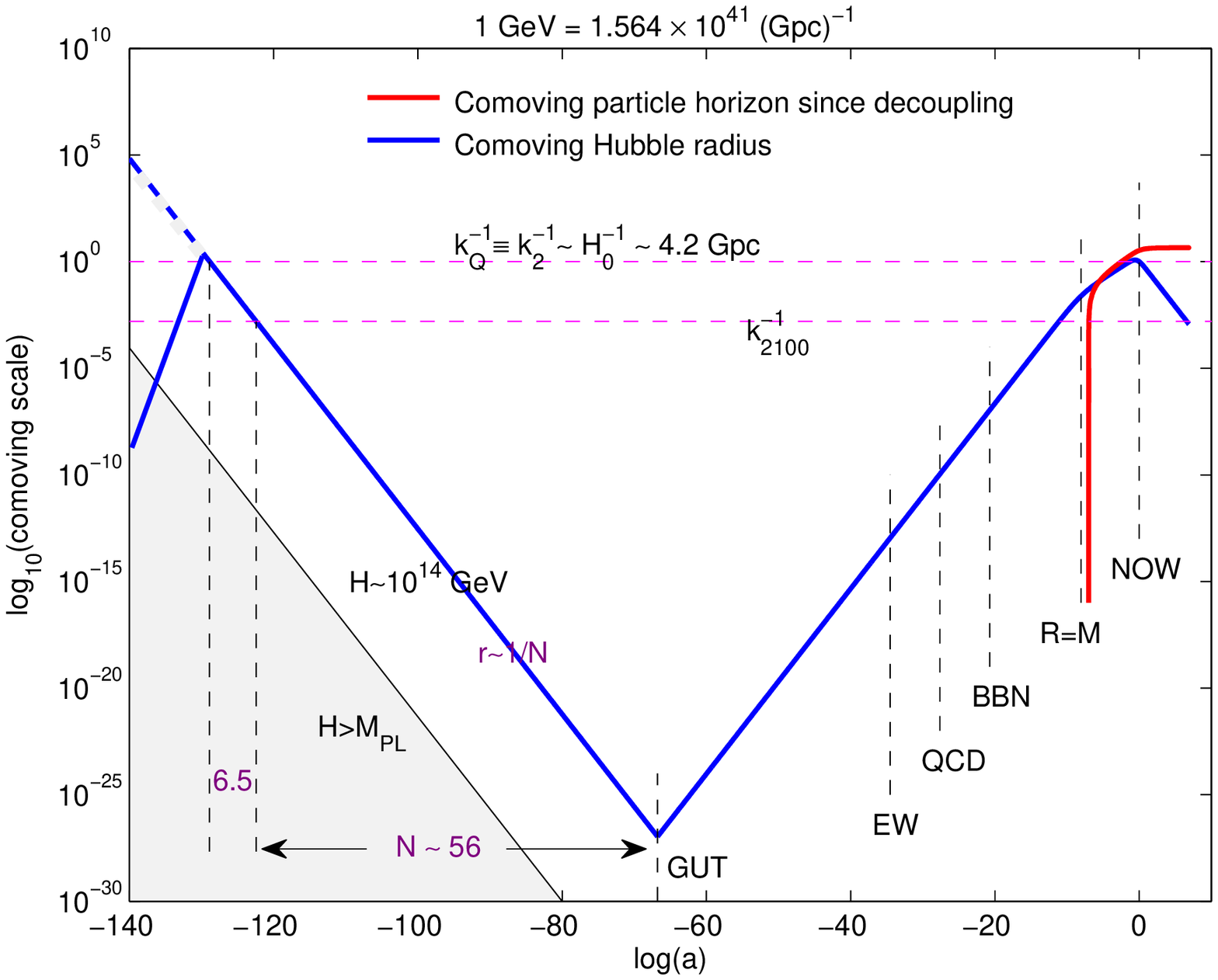}
\caption{The logarithm of  the comoving scales  and the logarithm
of  the comoving Hubble radius $ R_H =  1/[a \; H] $ vs. $ \log a $.}
\label{hr}
\end{figure}

It is very useful to plot the comoving scales of the cosmological fluctuation wavenumbers
and the comoving Hubble radius together [see fig. \ref{hr}]. One sees in this way how and when
the cosmological fluctuations cross out and in the Hubble radius.
The comoving Hubble radius is defined by $ R_H \equiv 1/[a(\tau) \; H(\tau)] $. We
display in Table VI the dependence of $ R_H $ on the scale factor $ a $ for all
the relevant eras of the universe.

\begin{table}[h]
  \begin{tabular}{|c|c|}
\hline  
Expansion stage & Dependence of $ R_H $ on $ a $\\
\hline \hline
Extreme Fast-roll &  $ a^2 $ \\ \hline
Fast-roll &  $ a^2/\sqrt{a^6 + {\rm constant}} $  \\ \hline
Slow--Roll inflation & $ 1/a $ \\ \hline
Radiation Dominated &  $ a $ \\ \hline
Matter Dominated &  $ \sqrt{a} $ \\ \hline
\end{tabular} 
\caption{Dependence of the comoving Hubble radius $ R_H =  1/[a \; H] $ 
on the scale factor $ a $ for the relevant eras of the universe.} 
\end{table}

The observed CMB quadrupole suppression can be easily explained 
if it exited the horizon by the end of fast-roll \cite{quadru,quamc}.
In that case, the modes which are horizon size today
had wavenumbers $ k_Q \simeq 11.5 \; m $ at horizon exit \cite{quamc}.
Combining this value of $ k_Q $ with the redshift since the pivot
wavenumber exited the horizon, eqs. (\ref{ka1}), (\ref{var}) and
(\ref{vk0}), determines the total redshift since the beginning of inflation to be
$$ 
z_{tot} = 0.9 \; 10^{56} \; \simeq e^{129} \; .
$$
Combining this value with the value of $ 1 + z_r \simeq 4 \; 10^{28} \simeq e^{66} $ 
by the end of inflation eq.(\ref{var}) yields a total number of $ N_{tot} = 63 $ inflation 
efolds. This value is very close to the minimal number of inflation efolds required 
to explain the entropy of the present universe due to photons and neutrinos 
\cite{biblia}:
$$
N_{tot} \geq 62.4 \; .
$$
Namely, this is the minimum number of inflation efolds compatible
with the present entropy of the universe.

In summary, assuming that the CMB quadrupole is suppressed because 
it exited the horizon by the end of fast-roll inflation {\bf fixes} 
the total number of inflation efolds which turns to be 
$$
 N_{tot} \simeq 63 \; .
$$

\begin{acknowledgments}
We thank Anthony Lasenby for fruitful discussions and his interest in this work.
\end{acknowledgments}

\end{document}